\documentclass[article]{aa} 
\usepackage{multirow} 
\usepackage{natbib}
\usepackage{amssymb,amsmath}
\usepackage{graphicx, subfigure}
\usepackage{siunitx}
\usepackage{url}
\usepackage{threeparttable}
\usepackage[table]{xcolor}
\usepackage{makecell}
\usepackage{epstopdf}
\usepackage{txfonts}
\usepackage{ textcomp }

\usepackage{color, colortbl}
\usepackage{booktabs}

\definecolor{Gray}{gray}{0.9}
\definecolor{Purple}{rgb}{0.88,0.5,1}
\definecolor{Orange}{rgb}{0.88,0.7,0.3}

\begin{document}

   \title{High-resolution observations of gas and dust around Mira using ALMA and SPHERE/ZIMPOL}

   \author{T. Khouri\inst{1}\thanks{{\it Send offprint requests to T. Khouri}\newline \email{theokhouri@gmail.com.}}, W. H. T. Vlemmings\inst{1},
   H. Olofsson\inst{1}, C. Ginski \inst{2}, E. De Beck \inst{1}, M. Maercker\inst{1}, S. Ramstedt\inst{3}.
}

\institute{Department of Space, Earth and Environment, Chalmers University of Technology, Onsala Space Observatory, 439 92 Onsala, Sweden 
\and
        Sterrewacht Leiden, P.O. Box 9513, Niels Bohrweg 2, 2300RA Leiden, The Netherlands 
        \and
Department of Physics and Astronomy, Uppsala University, Box 516, 751 20, Uppsala, Sweden
}

  \abstract
  {The outflows of oxygen-rich asymptotic giant branch (AGB) stars are thought to be driven by radiation pressure due to the scattering of photons
  on relatively large grains, with sizes of tenths of microns. The details of the formation of dust in the extended atmospheres of these stars
  and the mass-loss process is still not well understood.}
  {Constrain the distribution of the gas and the composition and properties of the dust grains that form in the inner circumstellar environment
  of the archetypal Mira variable $o$~Cet.}
  {We obtained quasi-simultaneous observations using ALMA and SPHERE/ZIMPOL on the Very Large Telescope (VLT) to probe the distribution
  of gas and large dust grains, respectively.}
  {The polarized light images show dust grains around Mira~A, the companion (Mira~B) and in a dust trail that connects the two sources.
  The ALMA observations show that dust around Mira~A is contained in a high-gas-density region
  with a significant fraction of the grains that produce the polarized light
  located at the edge of this region. Hydrodynamical and wind-driving models
show that dust grains form efficiently behind shock fronts caused by stellar pulsation or convective motions.
  The distance at which we observe the density decline (a few tens of au) is, however, significantly larger than expected for stellar-pulsation-induced shocks.
  Other possibilities for creating the high-gas-density region
  are a recent change in the mass-loss rate of Mira~A or interactions with Mira~B. We are not able to determine which of these scenarios is correct.
  We constrained the gas density, temperature, and velocity within a few stellar radii from
 the star by modelling the CO~$v=1, J=3-2$ line. {{ We find a mass $(\sim 3.8 \pm 1.3) \times 10^{-4}~M_\odot$ to be contained between the stellar
 millimetre  photosphere, $R^{\rm 338~GHz}_\star$, and $4~R^{\rm 338~GHz}_\star$.
 Our best-fit models with lower masses also reproduce the $^{13}$CO~$v=0, J=3-2$ line emission from this region well.
 We find TiO$_2$ and AlO abundances corresponding to 4.5\% and $< 0.1\%$ of the total titanium and aluminium expected for a gas with solar composition.}}
 The low abundance of AlO allows for a scenario
 in which Al depletion into dust happens already very close to the star, as expected from thermal dust emission observations and theoretical calculations
of Mira variables.
The expected relatively large abundance of aluminium
allows us to constrain the presence of aluminium oxide grains based on the scattered light observations
and on the gas densities we obtain. These models imply that aluminium oxide grains could account for
a significant fraction of the total aluminium atoms in this region only
if the grains have sizes $\lesssim 0.02~\mu$m. This is an order of magnitude smaller than the maximum sizes predicted by dust-formation and wind-driving
  models.}{ The study we present highlights the importance of coordinated observations using different instruments to advance our understanding of dust nucleation, dust growth, and wind driving in AGB stars.}
               
\titlerunning{Mira seen by ALMA and SPHERE}
\authorrunning{T. Khouri et al.}

\maketitle

\section{Introduction}

At the end of their lives, low- and intermediate-mass stars, $M \lesssim 8~M_\odot$, reach the asymptotic giant branch
(AGB) and experience strong mass loss. The mass-loss process is thought to be the result of the action of stellar pulsations
and strong convective motions that increase the density scale height of the atmosphere
and allow the formation of dust, followed by
radiation pressure acting on the { newly-formed grains} that drives the outflow \citep[e.g. ][]{Hoefner2018}.
The details of the wind-driving mechanism in oxygen-rich AGB stars (with a carbon-to-oxygen number ratio lower than one)
is particularly complex, because theoretical models
indicate that the dust grains need
to grow to tenths of micrometers in size to provide the required opacity through scattering \citep{Hofner2008}.
These large grains must have a low absorption opacity in order to remain relatively cold even close to the star
and, hence, not be destroyed \citep{Woitke2006}.
Observations have shown that such large grains indeed occur \citep[e.g. ][]{Norris2012,Ohnaka2016,Khouri2016}.

The sequence of dust species that condense in O-rich AGB stars and the composition of the grains that provide the scattering opacity
are still debated \citep[e.g. ][]{Gail2016,Gobrecht2016,Hoefner2016}.
Nonetheless, based on the elemental abundance of dust-forming elements, the condensation of some type of silicates is thought to be necessary
for the onset of the outflow. These silicates need to be iron-free in order to have low-absorption opacity \citep[e.g. ][]{Bladh2012}. Determining the composition of the grains seen in
scattered light is difficult, because the scattering opacities are virtually independent of grain composition. Nonetheless, interferometric
observations of dust emission and constraints from models of dust excess emission show that aluminium oxide grains exist very close
to the star \citep[at $\sim 2~R_\star$, e.g. ][]{Zhao-Geisler2012,Karovicova2013,Khouri2015}, while iron-bearing silicate grains are only seen
at larger radii \citep[at $\gtrsim 5~R_\star$, e.g. ][]{Zhao-Geisler2012,Karovicova2013}. The distribution of iron-free silicate grains is more difficult
to determine from observations of dust emission, because these grains have low absorption cross sections, and therefore remain relatively cold and
produce weak infrared emission.

{\bf To study the dust-formation and wind-driving processes and spatially resolve the wind-acceleration region,
we observed Mira { ($o$~Ceti)}, the archetypal Mira-type variable, using the
Atacama Large Millimetre Array (ALMA) and the Zurich imaging polarimeter (ZIMPOL) part of the spectro-polarimetric high-contrast exoplanet research (SPHERE) on the Very Large Telescope (VLT).}
The observations also resolve the stellar disc at visible and millimetre wavelengths.
The ZIMPOL and ALMA observations were acquired{ 18 days} apart,
and Mira was at post-minimum light phase ($\varphi \sim 0.7$). Mira{ AB} is a binary system consisting of an AGB star and a white-dwarf
or K dwarf \citep[e.g. ][]{Ireland2007} separated by roughly 90~au. The effect of the companion on the close environment of the AGB star is thought to be minimal
\citep{Mohamed2012}. On larger scales, Mira~B accretes material from the circumstellar envelope of Mira~A and the slow, dense outflow from Mira~A interacts
with a faster, but thin, outflow from Mira~B \citep[e.g. ][]{Ramstedt2014}.
Throughout this paper we adopt a distance of 102~pc to the Mira system, which is the weighted average between the distance obtained by Hipparcos
\citep[$92 \pm 10$~pc, ][]{vanLeeuwen2007} and from period-luminosity relations \citep[$110 \pm 9$~pc, ][]{Haniff1995}.
The stellar radius of Mira varies strongly with wavelength, as demonstrated by our observations and by data at other wavelengths
\citep[see e.g. the discussion in ][]{Kaminski2016}. For the discussion in this paper, we use the stellar radius measured from our ALMA observations,
 $\sim 21$~mas (see below), and a reference stellar radius of 15~mas obtained by \cite{Woodruff2009}
from measurements at 1.25~$\mu$m at post-minimum stellar pulsation phase, $\varphi = 0.7$. We refer to this stellar radius
measured in the infrared as $R^{\rm IR}_\star$.

\section{SPHERE observations}
\label{sec:obsSPHERE}
   
Mira was observed using filters NR, cnt748, and cnt820 by SPHERE/ZIMPOL on 27 November 2017 { (ESO programme ID 0100.D-0737, PI: Khouri)}.
The observations of the point-spread function (PSF) reference star (HD~12642) were carried out following the observations of Mira.
{ HD~12642 has a diameter of $\sim 2.3$~mas at visible wavelengths \citep{Bourges2017}, and is, therefore, a suitable PSF reference.}
The data were reduced using the (SPHERE/ZIMPOL) SZ software package developed at the ETH
(Eidgen\"ossische Technische Hochschule, Zurich). The basic procedures are standard \citep[bias frame subtraction, cosmic ray removal, and flat fielding; see e.g.][]{Schmid2017}
and essentially the same as those of the SPHERE reduction software provided by ESO.
The images of Stokes Q, U, and I were produced using the  expressions
\begin{equation*}
Q = \frac{Q_+ - Q_-}{2},~U = \frac{U_+ - U_-}{2}~~{\rm and}~~I = \frac{Q_+ + Q_- + U_+ + U_-}{4},
 \end{equation*}
 where $Q_+$, $Q_-$, $U_+$, and $U_-$ are the images obtained with the polarimeter
 oriented at 0$^\circ$, 45$^\circ$, 22.5$^\circ$, and 67.5$^\circ$, with respect to the north direction.
 { The polarized intensity and the polarization degree are calculated using
 \begin{equation}
 \label{equ:polInt}
 I^2_{\rm p} = Q^2 + U^2
 \end{equation}
 and
 \begin{equation}
 \label{equ:polDeg}
 p = I_{\rm p}/I,
\end{equation}
respectively.}
{ In Table~\ref{tab:SPHEREobs},
we present the range of values of airmass and seeing during the observations as provided by ESO in the header of the raw fits files
under the entries HIERARCH EZO TEL AIRM and HIERARCH ESO TEL IA FWHM, respectively.
We also give the detector integration time (DIT),
the number of exposures for a given orientation of the polarimeter in a given cycle (NDIT), and the total exposure time, which is equal to $4 \times {\rm DIT} \times {\rm NDIT} \times
{\rm NQU}$,
where NQU is the number of cycles for the acquisition of $Q_+, Q_-, U_+, {\rm and }~U_-$ images.}
{ The detector saturated in the observations of HD~12642 using filter NR. Hence we are not able to compare models to the data obtained using this filter,
as we do for filter cnt820 in Sect.~\ref{sec:modelDust}.}
The adaptive optics system of SPHERE
is expected to perform very well in the observational conditions and for bright sources such as Mira{ and HD~12642}.

\begin{figure}[t]
   \centering
      \includegraphics[width= 9cm]{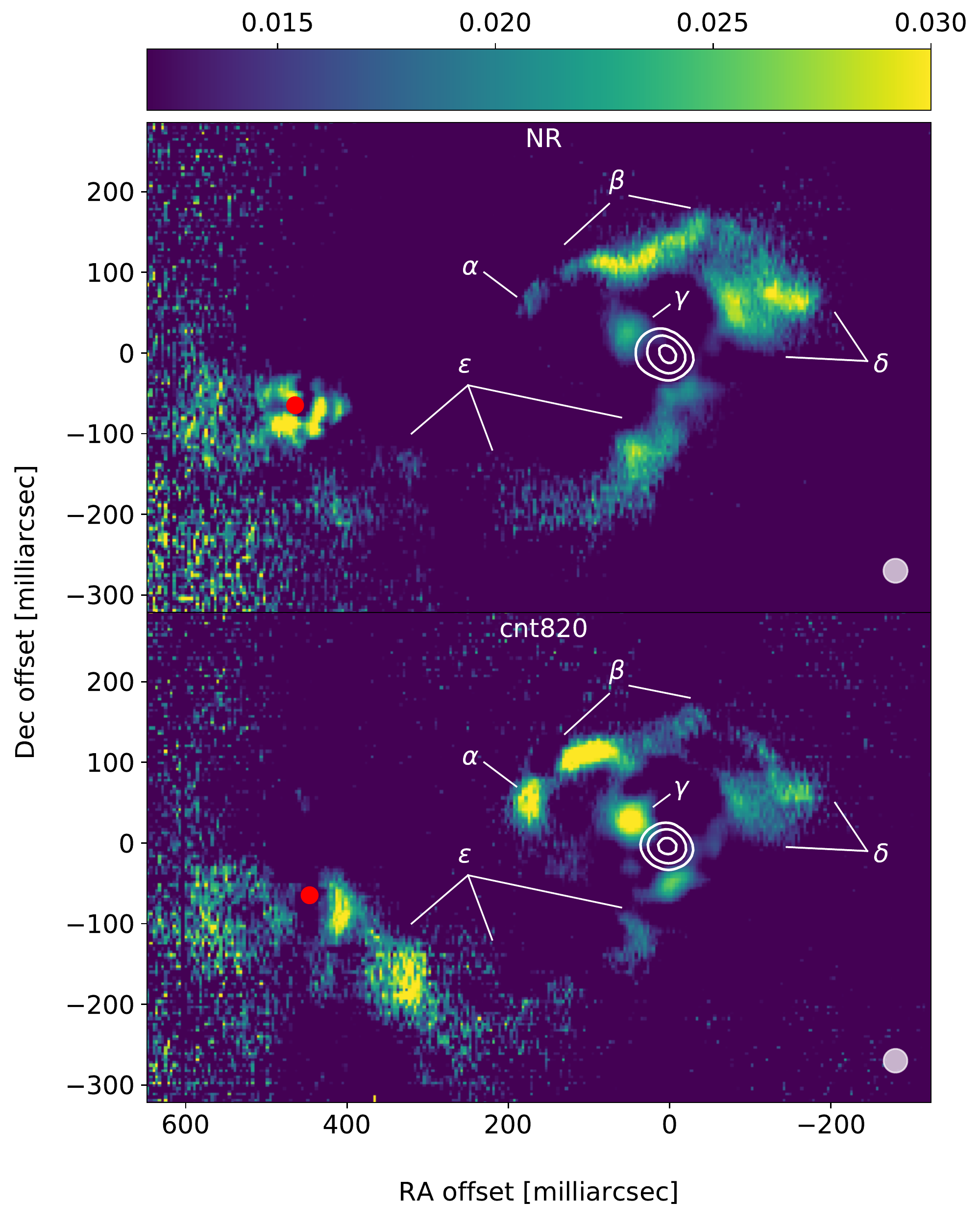}
      \caption{Polarization degree images obtained using ZIMPOL and filters NR ($\lambda_{\rm C} = 0.65~\mu$m) and cnt820 ($\lambda_{\rm C} = 0.82~\mu$m).
      The white contours show the total intensity
      at 80\%, 40\%, and 20\% of the peak value. The red circle marks the position of the companion and the white ellipses on the lower right corners show
      the FWHM of the observed PSF reference star at each given filter. { The offsets in declination and right ascension are given with respect to the position of Mira~A.}}
         \label{fig:polDegNR}
   \end{figure}

In the total intensity images,{ Mira~A and Mira~B} can be seen in all filters,
with the intensity of the compact companion decreasing relatively to that of the primary for longer wavelengths.
In the images, the full-width at half maximum (FWHM) of the point-spread-function (PSF) reference star is 29.9~mas at $0.65~\mu$m { (with a small degree of saturation)}
and 29.5~mas at both $0.75~\mu$m and $0.82~\mu$m.
The FWHM of Mira~A is 42.1~mas at $0.65~\mu$m, 43.9~mas at $0.75~\mu$m, and 41.8~mas at $0.82~\mu$m.
The stellar disc of Mira~A is, therefore, marginally resolved in the observations. We find Mira~A to be asymmetric at
$0.65~\mu$m and $0.75~\mu$m, extending in the north-east direction (see Fig.~\ref{fig:polDegNR}).
At $0.82~\mu$m, the stellar disc of Mira~A does not appear significantly asymmetric.

The polarized intensity was computed from the observed images of the Stokes parameters Q and U (these images are shown in Fig.~\ref{fig:Q+U_images} in
the Appendix). The data reveal structures whose relative strengths change considerably between filters.
This can be seen in the images of the polarization degree shown in Fig.~\ref{fig:polDegNR} for filters NR and cnt820.
We do not discuss the observations using filter cnt748, because the images are consistent with
those in filters NR and cnt820, but the signal-to-noise ratio in the images of polarized light is significantly lower.
We have labelled the different dust structures observed in Fig.~\ref{fig:polDegNR} to facilitate the discussion.
{ Based on the spread of the values obtained for the individual exposures,
the uncertainties on the reported polarization degree per pixel are ${\sim 0.02}$ at the brightest regions of structures $\alpha$ and $\beta$,
${\sim  0.01}$ at structure $\gamma$, and ${< 0.01}$ closer to the central star.
We note that the uncertainty on the average polarization degree of a given structure (which encompasses several pixels) would be significantly smaller
than the values given above.
The offsets in right ascension and declination given throughout the paper are with respect to the position of Mira~A.}

We detect polarized light close to Mira~A, as expected, but also from around Mira~B and from what seems to be a dust trail connecting the two stars.
The degree of polarization peaks at ${\sim  3.2}$~\% in NR, ${\sim 3.5}$~\% in cnt748, and ${\sim 4.0}$~\% in cnt820.
These are relatively low values when compared to other oxygen-rich AGB
stars observed using ZIMPOL.{ This is particularly true when one considers
that Mira~A was at post-minimum light phase at the time of the ZIMPOL observations and observations of other close-by AGB stars were obtained
at a similar light phase.}
For instance, ZIMPOL observations of W~Hya
revealed a polarization degree peaking at 18\% at minimum light phase and $\sim 13\%$ at pre-maximum light phase \citep{Ohnaka2017}, while
observations of R~Dor also show the peak of the polarization degree varying between $\sim 7\%$ and $\sim 14\%$
at two different epochs \citep{Khouri2016}. Moreover, ZIMPOL monitoring of R~Dor with observations at 11 epochs over eight months show the polarization
degree to reach peak values $> 7\%$ at all epochs (Khouri et al., in prep.).
The polarization degree also peaks at larger distances from the star for Mira~A ($> 5~R_\star$) than for W~Hya and R~Dor (${\sim 2~R_\star}$).
One important difference between Mira~A and the other two stars is that Mira~A is a Mira-type variable, while W~Hya and R~Dor are classified as semi-regular pulsators.

 \begin{table}
\footnotesize
\setlength{\tabcolsep}{12pt}
\caption{Log of the SPHERE observations.}              
\label{tab:SPHEREobs}      
\centering                                      
\begin{tabular}{l r l}
\hline
\hline
{ NR} \\
Mira \\
& Airmass & 1.077 -- 1.083 \\
& Seeing & 0.62 -- 0.87 arcsec\\
& ND filter & None \\
& DIT & 1.25 \\
& NDIT & 18 \\
& Exp. time & 27 min\\
HD~12642 \\ 
& Airmass & 1.141 -- 1.159 \\
& Seeing & 0.76 -- 0.83 \\
& ND filter & ND\_1\\
& DIT & 5.0 \\
& NDIT & 8 \\
& Exp. time & 16 min\\
{ cnt748 + cnt820} \\
Mira \\
& Airmass & 1.083 -- 1.111 \\
& Seeing & 0.67 -- 0.78 \\
& ND filter & ND\_1 \\
& DIT & 1.6 \\
& NDIT & 20 \\
& Exp. time & 25.6 min\\
HD~12642 \\
& Airmass & 1.162 -- 1.188 \\
& Seeing & 0.73 -- 0.83 \\
& ND filter & ND\_1\\
& DIT & 5.0 \\
& NDIT & 10 \\
& Exp. time & 10 min\\

\hline
\end{tabular}
\end{table}

\subsection*{Beam-shift effect}
\label{sec:beam-shift}

The ZIMPOL observations of Mira show signs of being affected by a beam-shift effect. This is a known problem that is caused
by sub-pixel shifts between the images obtained at different orientations of the polarimeter \citep{Schmid2018}.
This leads to one negative and one positive lobe that appear in
opposition in the images of the Stokes parameters Q and U, as can be seen in our data, particularly in the cnt820 images (see Fig.~\ref{fig:Q+U_images} in the Appendix).
This beam-shift effect causes residuals to appear very close to the central star after polarimetric differential imaging. These residuals can be mistaken for a polarization signal,
and need to be quantified when analysing polarized data in the innermost region.

We have examined the beam-shift effect in the cnt820 images.
We find that a shift of $\lesssim 0.05$~pixels { (or 0.18~mas)}  in the north-east--south-west direction is needed to reproduce the lobes in the
obtained Q images in filter cnt820, while a shift of similar magnitude roughly in the opposite direction reproduces the lobes seen in the U images
in the same filter. { These results are only valid for the observations we present since the beam-shift effect is highly dependent on many parameters
\citep[see discussion in ][]{Schmid2018}.
This shift produces artificial polarized light
that reaches polarization degrees of $\lessapprox 2\%$ to the north-east and to the south-west of the central star up to $\sim 70$~mas from the centre.
The polarization vectors produced by the beam-shift effect, in the specific case of the images in filter cnt820, are aligned (with respect to Mira~A) tangentially
in the north-east region and radially in the south-west region.
Since the polarization vectors of single-scattered light from a central source are expected to be tangentially aligned, this would lead to constructive and
destructive interactions with the polarized signal in the north-east and south-west, respectively. In this way, the polarization degree of Clump~$\gamma$ (see Fig.~\ref{fig:polDegNR})
is probably overestimated, { especially} in the cnt820 images. Moreover, the vectors of the polarized light we observe close to { Mira~A} are aligned tangentially to the star
in the north-east, but radially in the east (see Fig.~\ref{fig:polVectors}). { This is consistent with our estimation of the beam-shift effect in the cnt820 images.}

{ Scattering of radiation by grains with sizes comparable to the wavelength of the incident radiation divided by $2 \pi$ can cause polarization inversions for certain
scattering angles \citep[see e.g.][]{Aronson2017}. Such polarization inversions, or the presence of aligned dust grains \citep[such as observed in e.g. VY~CMa, ][]{Vlemmings2017a},
could in principle be invoked to explain the radially aligned polarization vectors. The fact that we are able to reproduce the observed orientation and strength of the
polarization vectors by introducing a beam-shift of the order of that seen in the Q and U images (Fig.~\ref{fig:Q+U_images}) shows that no astrophysical explanation
needs to be
invoked for the radially-aligned polarization vectors in the images we present.

The beam-shift effect appears to have a different outcome close to Mira~B compared to Mira~A at 0.82~$\mu$m (Fig.~\ref{fig:polVectors}),
although the same pattern and intensity of beam-shift effect is expected
for both sources in the same image.
We find two reasons for the observed difference. First, 
Mira~B is relatively weak at 0.82~$\mu$m. Therefore, the polarization due to the beam-shift effect around Mira~B is relatively weaker than around Mira~A.
This leads to polarization due to scattering {\bf of radiation off} dust grains to dominate close to Mira~B
because the radiation field is dominated by Mira~A.
Second, at 0.82~$\mu$m the speckle ring at the adaptive-optics control radius \citep{Schmid2018}
appears on top of Mira~B and significantly affects the total intensity and the polarization degree images.
The brightest speckle at the adaptive-optics control radius can be seen in Fig.~\ref{fig:polVectors}
as a dark region to the north of Mira~B at 0.82~$\mu$m and to the west at 0.65~$\mu$m.}

From our estimates of the residual polarization from the beam-shift effect and the polarization degree in the obtained images,
we conclude that the intrinsic polarization degree is $\lessapprox 2\%$ in the innermost region around Mira~A ($r \lessapprox 84$~mas).
The beam-shift effect could in principle be removed by improving the alignment of the images during the reduction process \citep{Schmid2018}.
However, since the effect is relatively small in the data we report
and we can set a meaningful upper limit to the polarization degree close to the star, we do not attempt to improve the reduction.  

    \begin{figure}[t]
   \centering
      \includegraphics[width= 9cm]{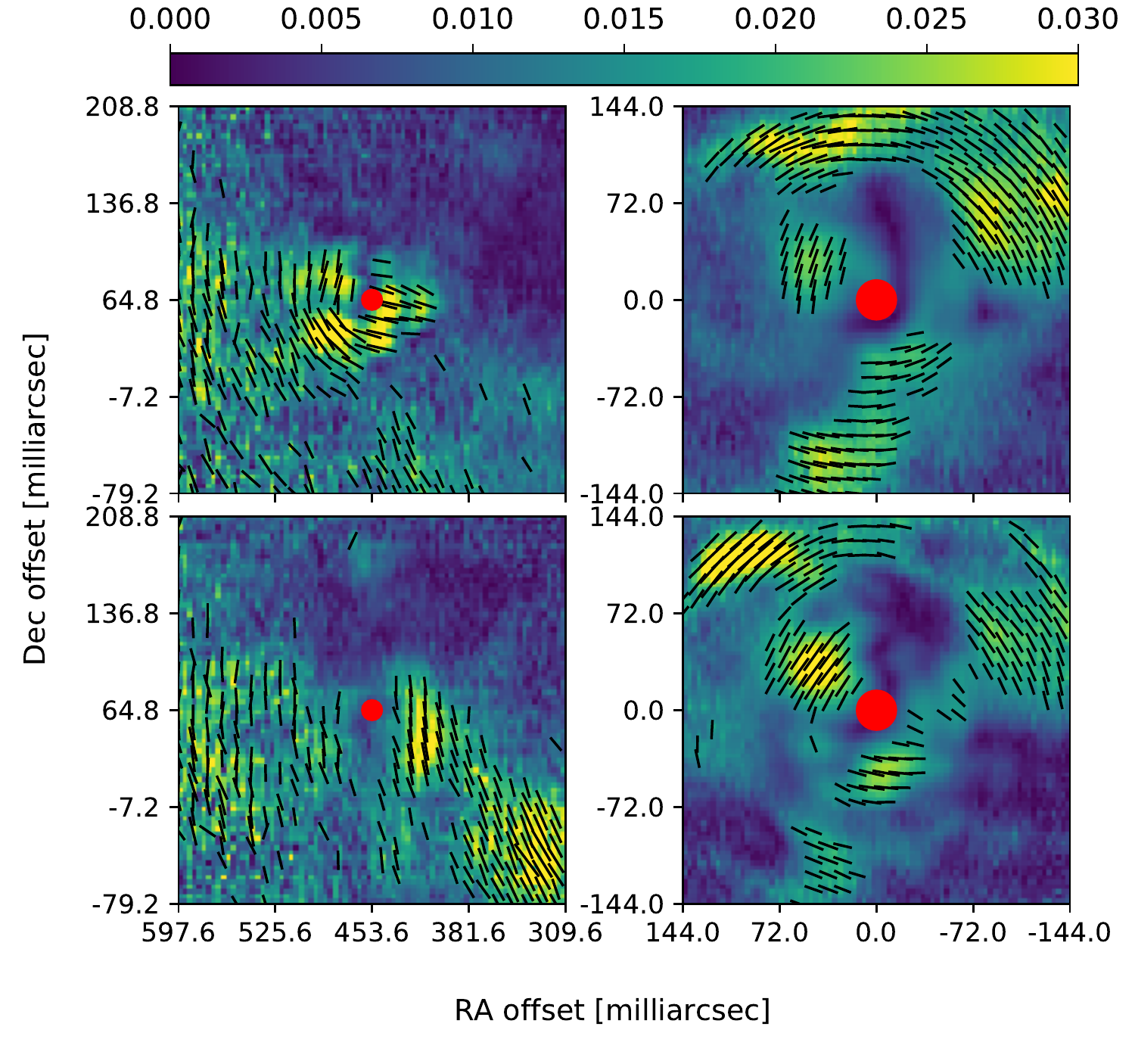}
      \caption{Polarization degree (shown by the colour map) and direction of the polarization vectors around Mira~A (right panels) and Mira~B (left panels).
      The upper and lower panels show the image in the NR and the cnt820 filters, respectively. The red dots mark the positions
      of the two stars. The vectors are drawn when the polarization degree is higher than 1.5\%. { The offsets in declination and right ascension are given with respect to the position of Mira~A.}}
         \label{fig:polVectors}
   \end{figure}

\section{ALMA observations}

The ALMA observations were obtained on 9 November 2017 (project 2017.1.00191.S, PI: Khouri), about 18 days before the ZIMPOL observations.
The observations were performed using four spectral windows
(spws) of 1920 channels each. The velocity resolution was
$\sim 0.85$~km~s$^{-1}$ and the spws
covered 329.25 -- 331.1~GHz, 331.1 -- 333.0~GHz, 341.35 -- 343.25, and 343.2 -- 345.1~GHz.
The full calibration and imaging was performed in CASA
5.1.1. The initial gain calibration (bandpass, amplitude, absolute flux-density scale, and phase) was performed following the
standard ALMA procedures.
{ Observations of J0237+2848 were used for flux and bandpass calibrations and of J0217+0144 for gain calibration.}
We then imaged the line free channels and performed
a self-calibration step using the stellar continuum.
These phase solutions were also applied to the full dataset.
The final image products were created using Briggs robust weighting{ (with robust parameter equal to 0.5)}, resulting in a beam with
major and minor axes of 32.7~mas and 19.1~mas, respectively, and a position angle of 53.2 degrees.
{Root-mean squared (rms) noise levels of $\sim 1.4$~mJy/beam and $\sim 2.0$~mJy/beam were achieved, respectively, in the final images of the higher and
lower frequency spectral windows} for a spectral resolution of $\sim 1.7$~km~s$^{-1}$.
In the continuum images, we find an rms noise of 0.04 mJy/beam.

By performing a fit of a uniform-disc source to the observed visibilities using the {\it uvmultifit} code \citep{Marti-Vidal2014},
we determine the size of Mira~A to be $(41.9 \pm 0.1) \times (42.5 \pm 0.1)$~mas (at PA -$83 \pm 3$ degrees) with a brightness temperature of 2100 K. Therefore,
the stellar disc of Mira~A is seen to be roughly circular at $\sim 338$~GHz at the time of the ALMA observations.{ We find a continuum flux density for Mira~A of $250.6 \pm 0.2$~mJy.
The residuals obtained after the subtraction of our uniform-disc model have maximum and minimum values of 1.9 and -2.8~mJy, respectively.}
We refer to the stellar radius at 338~GHz of 21.2~mas as $R^{\rm 338 GHz}_\star$, that is $R^{\rm 338 GHz}_\star = 1.4~R^{\rm IR}_\star$.

 \begin{figure}[t]
   \centering
      \includegraphics[width= 8cm]{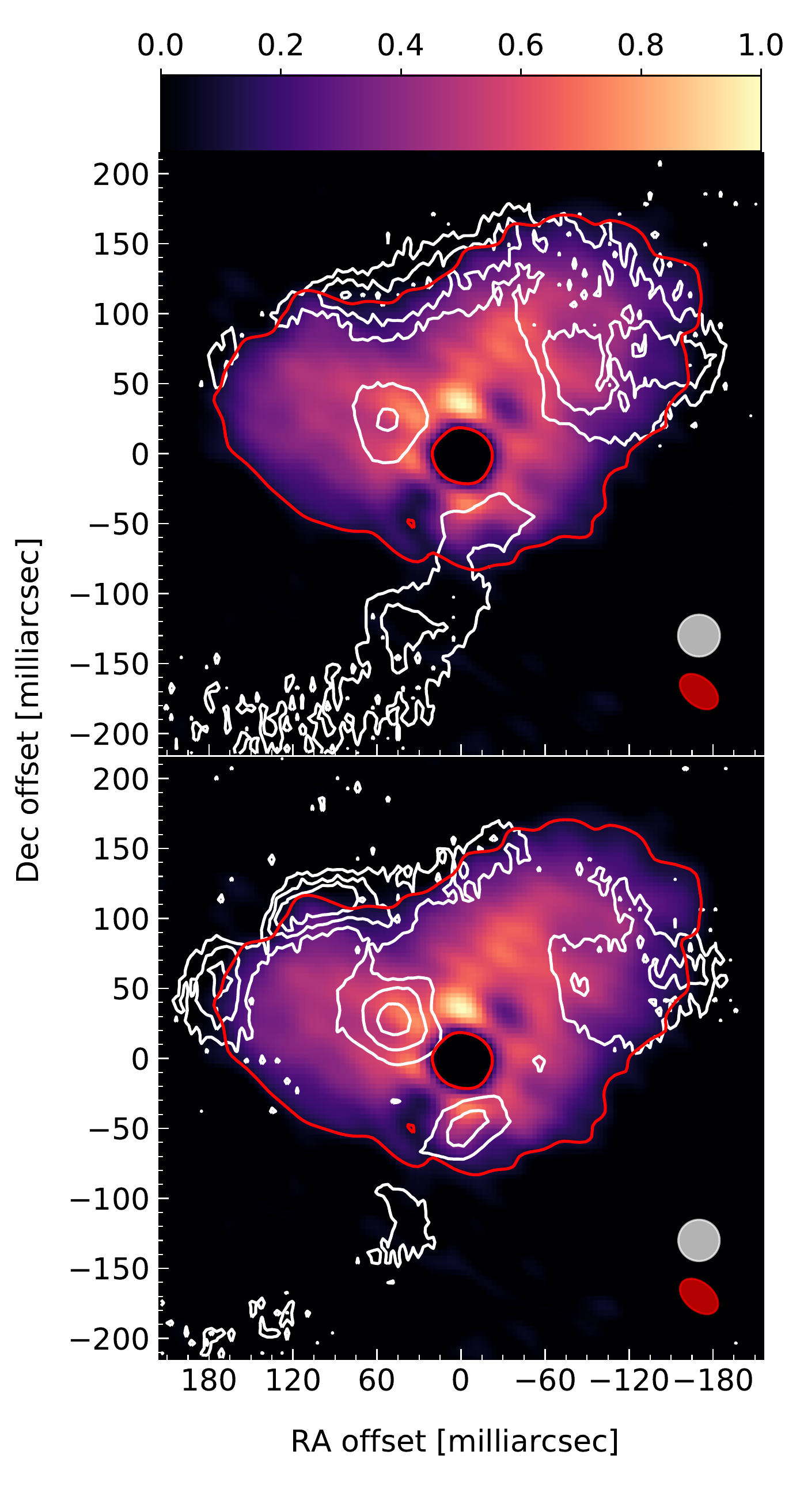}
      \caption{Comparison between a low-excitation SO~$N_J=8_8 - 7_7$ line observed using ALMA
      and the polarized light observed using ZIMPOL.
      The SO $N_J=8_8 - 7_7$ integrated line intensity was normalized to the peak value in the image and is shown by the colour scale.
      The red contour shows the 10\% level of the SO emission, and represents the edge of the emission region. The white contours 
      show the 1.5\%, 2.3\%, and 3.1\% levels of the polarization degree
      observed using ZIMPOL and filters NR, $\lambda_{\rm C} = 0.65~\mu$m (top), and cnt820, $\lambda_{\rm C} = 0.82~\mu$m (bottom).
      The white and red ellipses on the lower right corner show, respectively,
      the FWHM of the observed PSF reference star at the given filter in the visible and the beam of the ALMA observations.
      { The offsets in declination and right ascension are given with respect to the position of Mira~A.}}
               \label{fig:polDeg+SO}
   \end{figure}

A spectrum extracted within an aperture of 50~mas in radius centred on the star reveals more than a 100 spectral lines
of which approximately one third (consisting mainly of weak lines) has not yet been identified by us.
The identified lines correspond to transitions from CO, $^{13}$CO, C$^{18}$O,
SiO, $^{29}$SiO, $^{30}$SiO, Si$^{17}$O, $^{29}$Si$^{17}$O, SO, $^{33}$SO, SO$_2$,
TiO$_2$, AlO, PO, and H$_2$O.
In this study, we focus on compact emission revealed by strong low-excitation lines of molecules such as SO and $^{13}$CO, on the
vibrationally-excited ($v=1$) rotational line CO~$J=3-2$ at 342.647~GHz \citep{Gendriesch2009},
and on lines of molecules relevant for dust formation (TiO$_2$ and AlO)
in comparison to the images of polarized light. A comprehensive overview of the ALMA
observations will be presented in a future paper. 
All the images of spectral line emission presented by us are stellar-continuum subtracted.
The dark region at the stellar position is caused by{ molecular} absorption against the stellar continuum.

In the ALMA data obtained by us,
emission from transitions with upper-level excitation energy $< 200$~K arises from a region extending mostly in the northern
hemisphere and reaching distances of up to $\sim 200$~mas
in radius from the centre of Mira~A (see Fig.~\ref{fig:polDeg+SO}).
Two lobes can be identified in this emission region (see Fig.~\ref{fig:SO-channels} in the Appendix), one extending to the east and blue-shifted
and another extending to the north-west and slightly red-shifted with
respect to the local standard of rest velocity of the centre of mass of the system, $\upsilon_{\rm LSR} \sim 47$~km/s
\citep[e.g.][]{Kaminski2017}.
{ Throughout this paper, we refer to this region (delimited by the red line in Fig.~\ref{fig:polDeg+SO}) as the compact molecular line emission region.}
As we argue in Section~\ref{sec:gas_x_dust}, this emission region is probably
defined by a{ steep gas density decline} at its outer edge.

  \begin{figure}[t]
   \centering
      \includegraphics[width= 9cm]{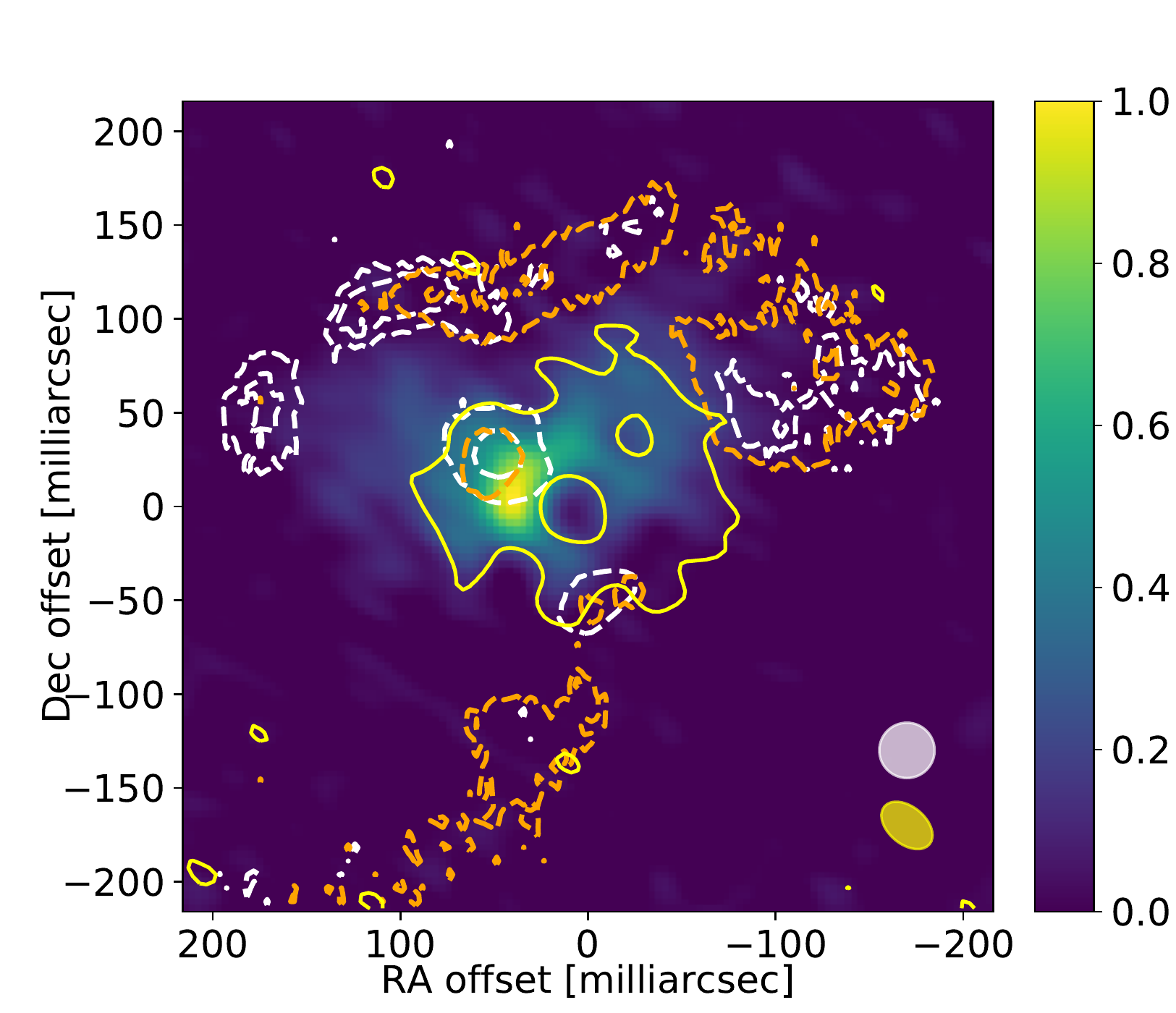}
      \caption{Comparison between the CO~$v=1, J=3-2$ and stacked TiO$_2$ lines observed using ALMA and the polarized light observed using ZIMPOL.
      The colour map shows the stacked emission from the nine unblended TiO$_2$ lines (see Table~\ref{tab:obs}) normalized to the peak value.
      The 2\% and 3\% levels of the polarization degree at 0.65~$\mu$m and 0.82~$\mu$m are shown by the dashed white and orange lines, respectively.
      The yellow solid contour shows the 10\%
      level of the CO~$v=1, J=3-2$ emission. The white and yellow ellipses on the lower right corner show, respectively,
      the FWHM of the observed PSF reference star in the NR filter (which is very similar to that using filter cnt820) and the beam of the ALMA observations.
      { The offsets in declination and right ascension are given with respect to the position of Mira~A.}}
         \label{fig:clumpB}
   \end{figure}

\subsection*{Maximum recoverable scale}
\label{sec:MaxScale}

The lines we discuss in this study are $^{13}$CO~${J=3-2}$, AlO~${N=9-8}$, SO~${N_J=8_8 - 7_7}$, CO~${v=1, J=3-2}$, and several lines of TiO$_2$.
Among these, the emission regions of the $^{13}$CO and SO lines are expected to be more extended than the compact molecular line emission region we find.
Such extended
emission would be filtered out by the ALMA observations we report, which have a maximum recoverable scale of ${\sim 400}$~mas. The maximum extent of the structure
we observe is $< 300$~mas in the moment zero maps and $\lessapprox~200$~mas in the images of individual channels. This is
significantly smaller than the maximum recoverable scale. The flux from lines with emission regions expected to be
smaller than the maximum recoverable scale should be almost completely recovered.  This is the case for AlO \citep{Kaminski2016},
CO~$v=1$ \citep{Khouri2016b}, and TiO$_2$ \citep{Kaminski2017}.
To confirm this expectation, we performed simulations of observations
of uniform discs by the array configuration used to observe Mira. We find that for a uniform disc with a diameter of 300~mas about 20\% of the flux is
lost when the visibilities are re-imaged, while for a 200~mas uniform disc model, only 5\% of the flux is lost.

The $J=3-2$ line of $^{13}$CO has an emission region much larger than the scales probed by the observations we report.
Therefore, a large fraction of the smooth, large-scale emission
is expected to be resolved out. This also affects the flux we estimate from the innermost region.
To quantify how much flux density per beam is resolved out because of large-scale emission,
we compared the flux we measure from the compact molecular line emission region to the observations
obtained by \cite{Ramstedt2014}.
{ We imaged the $^{13}$CO~$J=3-2$ line (PI: Ramstedt) using Briggs robust weighting (with robust parameter equal to 0.5) and obtained images
with a beam of $0.68 \times 0.45$~arcsec, and an area of 0.24~arcsec$^{2}$. The compact molecular line emission region (with an area of $0.046$~arcsec$^2$) is
thus unresolved in these earlier observations.
We find that the compact molecular line emission region accounts for 72\% of the
flux density in the central beam of the lower resolution image, while encompassing only 19\% of the total area.
Assuming that the resolved-out emission is smooth and that, therefore, the remaining 28\% flux density is spread
evenly over the area of 0.24~arcsec$^{2}$, we find that $\sim 5\%$ ($28\% \times 0.046 / 0.24$) 
of the flux density in the central beam of the lower resolution observations
is not recovered in the compact molecular line emission region.
Therefore, the flux density measured in the compact molecular line emission region should have been
about 7\% larger. This emission would have been lost in our observations due to the filtering of large-scale emission.}

\section{Results}

\subsection{The distribution of gas and dust}
\label{sec:gas_x_dust}

First,{ we discuss the nature of the compact molecular line emission region seen in the ALMA images.
We consider that the observed compact emission is most likely
the consequence of a steep gas density decline at its outer edge.
An alternative explanation could be that the low-excitation transitions are efficiently radiatively excited in this region or that there is a sudden change of
the gas temperature at the region's outer edge.
We consider these possibilities very unlikely, because
lines of different molecules (SO, SO$_2$, $^{29}$SiO, $^{13}$CO, AlO, TiO$_2$) with different upper-level energies
show a very similar emission region. Moreover,
the corresponding upper levels of the transitions have relatively low excitation energies and 
the relatively high gas densities we find ($\gtrsim 10^8$~cm$^{-3}$, see Sect.~\ref{sec:columnDens}) imply that these low-excitation transitions should be 
efficiently excited by collisions.}

{ A clear correlation between the distributions of gas and dust can be seen when the ALMA and SPHERE data are compared.}
In Fig.~\ref{fig:polDeg+SO}, we show the moment zero map of the SO line and the polarization degrees at $0.65~\mu$m and $0.82~\mu$m.
While Clump~$\alpha$ and Structure~$\beta$ follow the edge of the compact molecular line emission region, Structure~$\delta$ appears on
top of the north-western lobe. The observed structures and the varying strength of polarized intensity is probably partially caused
by different scattering angles in a somewhat hollow dust shell.
In this case, the angle-dependence of the phase function and polarization efficiency can make the polarization degree vary significantly as a function of wavelength.
Grain size variations between the different structures might also contribute to the observed differences in polarization degree. 
Since we do not know the geometry of this region along the line of sight, a detailed interpretation{ of the cause of the varying polarization degree} is difficult.

{ Second, we discuss the possible causes for the existence of the compact molecular line emission region and the correlation between the distribution of gas and dust.
Hydrodynamical models show that the propagation of a shock
\citep[e.g. ][]{Hofner2003,Ireland2011,Freytag2017} produces a density contrast at the shock front and
causes dust formation to happen efficiently in the post-shocked gas \citep{Freytag2008}.
These two features qualitatively match our observations but
stellar-pulsation-induced shocks are expected to be important only very close to the star \citep[up to a few stellar radii, ][]{Hoefner2018} before dissipating.
Since the edge of the compact molecular line emission region extends up to $\sim 12~R^{\rm IR}_\star$,
we consider it unlikely that the compact molecular line emission region is the consequence of pulsation-induced shocks.}

{ Another} possibility is that the mass-loss rate of Mira increased significantly recently. In this case, the edge of the molecular emission region would be
the consequence of the density contrast between a previous lower mass-loss rate wind and a recent higher mass-loss rate wind.
Interestingly, an X-ray outburst was seen towards Mira~A
by observations made in 2003 \citep{Karovska2005}. The authors speculated
that the burst could have significant consequences for the mass-loss rate of Mira~A.
To cover a distance of $\sim 18$~au, to the edge of the compact molecular line emission region, over the period of 14~years that
separates the observations reported by us from those by \cite{Karovska2005}, gas would need to travel at a speed of $\sim 6$~km/s.
This expansion speed is within the range expected for the inner regions of an AGB outflow and thus this is a plausible scenario.
Finally, another possibility is that
this high density region is caused by interactions with the binary companion{, but the effect of the companion
on gas very close to Mira~A is expected to be small \citep{Mohamed2012}}. Future observations that reveal the evolution of the system will
be able to distinguish between these different possibilities.

{\renewcommand{\arraystretch}{1.3}
 \begin{table*}
\footnotesize
\setlength{\tabcolsep}{10pt}
\caption{{ Parameters of the best-fit model to the observed CO~$v=1, J=3-2$ line. We also give the preferred parameters when also including the $^{13}$CO~$J=3-2$ line
and the range over which parameters were varied.} }             
\label{tab:vibCO-model}      
\centering                                      
\begin{tabular}{l@{} r l l r@{\phantom{ai}}p{10pt}@{} l l }
\multicolumn{2}{c}{Parameter} & \multicolumn{2}{c}{Best models} & & Grid &  & Constraints\\
\multicolumn{2}{c}{} & $^{12}$CO & Incl. $^{13}$CO & & range &  & \\
\hline
\hline
$d$ & [pc] & 102 & 102 & &--& & Assumed \\ 
$T_\star$ & [K] & 2100 & 2100 & &--& & Spatially-resolved continuum\\
$R^{\rm 338~GHz}_\star$ & [au] & 2.15 & 2.15 & &--& & Spatially-resolved continuum\\
$R_{\rm out}$ & [au] & 8.6 & 8.6 & &--& & Emission region \\
$T_\circ$ & [K]& $900^{+50}_{-150}$ & 900 & 500 &--& 1700 & Fit + $T_{\rm bright}$ optically-thick lines\\
$\epsilon$ & & $0.35^{+0.05}_{-0.3}$ & 0.24 & 0.0 &--& 0.9 & Fit\\
$\upsilon_{\rm sto}$ & [km/s] & $4.0 \pm 0.3$ & 4.0 & 3.0 &--& 5.0 & Fit \\ 
$\upsilon_{\circ}$ & [km/s] & $-12 \pm 2$  & -12 & -18 &--& -9 & Fit \\ 
$\upsilon_{\rm f}$ & [km/s] & $5 \pm 1$ & 5 & 3&--&10 & Fit \\ 
$\beta$ & & $0.4 \pm 0.1$ & 0.4 & 0.1&--&0.7 & Fit \\
$n_{\circ}$ & [cm$^{-3}$]& $(2.5^{+0.8}_{-0.5}) \times 10^{11}$\phantom{~~~} & $2.5 \times 10^{11}$ & $1 \times 10^{11}$&--&$7\times10^{11}$ & Fit \\
$\eta$ & & 2.25$^{+0.75}_{-0.25}$ & 2.75 & 1.0&--&4.0 & Fit \\
$\upsilon_{\rm LSR}$ & [km/s] & 47.7 & 47.7 & &--& & Fit \\
\hline
\end{tabular}
\end{table*}}

Dust Clump~$\gamma$ is the one located closest to Mira~A
in the polarized-light images,{ located at a distance of $\sim 3.5~R^{\rm IR}_\star$ from the centre of Mira~A}.
As discussed in Sect.~\ref{sec:beam-shift}, the polarized light detected in Clump~$\gamma$ is probably
at least partially caused by the beam-shift effect. Nonetheless, the existence of this clump seems real, since it is seen
in images in both filters. The amount of polarized light that originates from scattering {\bf of radiation on} dust grains is, however, somewhat uncertain.
The location of Clump~$\gamma$ is probably not only a projection effect because polarization through scattering
would not be efficient if it were located too far out of the plane of the sky.
Indeed, the image of the vibrationally-excited CO~$J=3-2$ line shows that Clump~$\gamma$
is most likely embedded in a region of very high gas density (see Fig.~\ref{fig:clumpB}),
with number density $> 10^{10}$~cm$^{-3}$  (see below).
Interestingly, the stellar disc seems to extend in the direction of Clump~$\gamma$
at $0.65~\mu$m (see Fig.~\ref{fig:polDegNR}) and $0.75~\mu$m.
{ A similar coincidence between an extension of the stellar photosphere and a dust clump has been reported for Betelgeuse \citep{Kervella2018},
which was interpreted to be
caused by preferential mass loss from the pole of the star.}

We note that the stellar continuum of Mira~A measured by ALMA does not appear to be asymmetric.
Therefore, while the images in visible light at 0.65~$\mu$m and 0.75~$\mu$m show an extension of Mira~A toward Clump~$\gamma$,
those at 0.82~$\mu$m and in the sub-millimetre do not.
Molecular opacity is expected to affect
the observations at 0.65~$\mu$m and 0.75~$\mu$m more strongly than those at 0.82~$\mu$m, although the difference in source size between images at
0.82~$\mu$m and the other filters is small in the ZIMPOL
observations (see Sect.~\ref{sec:obsSPHERE}). In the sub-millimetre, free electrons rather than molecules dominate the continuum opacity. Hence
a possible explanation for Mira~A appearing extended only at 0.65~$\mu$m and 0.75~$\mu$m is that molecular opacity is large enough
locally to affect the appearance of Mira~A at these wavelengths but not enough to affect its appearance at 0.82~$\mu$m. In this scenario,
the sub-millimetre opacity due to free
electrons would also need to be small enough not to produce an effect on the 338~GHz continuum.
This suggestion is speculative, given the data at hand, and more observations are needed to further investigate the cause of asymmetries of this type.

\subsection{Dust around Mira~B and the dust trail}

The dusty Trail~$\epsilon$ marked in Fig.~\ref{fig:polDegNR} seems to connect the two stars. It is not obvious from the ZIMPOL images whether this traces an accretion
flow to Mira~B, the interaction between the outflows of the two stars, or some other process. We note that the ALMA observations do not reveal any structure
related to this in the many spectral lines
observed. A different bridge-like component was seen
in X-rays using {\it Chandra} and in {\bf ultraviolet light} using the { Hubble Space Telescope}
\citep{Karovska2005}. While the dust trail we report bends through the southern hemisphere, the bridge in the
X-rays connects the two sources through a roughly straight line. It is not clear whether this is because of time variability
of the gas and dust density distribution or if these different observations trace distinct and different structures.

The polarized light detected towards the companion at 0.65~$\mu$m and 0.82~$\mu$m is most likely produced by scattering of radiation off dust grains.
 \cite{Ireland2007} found evidence of a disc surrounding Mira~B, with an outer radius of $\sim 10$~au, using observations of thermal dust emission at wavelengths
 between $8$ and $18~\mu$m. The scattered light we see might be produced by the structure identified by them.
Although Mira~A outshines Mira~B at visible wavelengths,
radiation from the compact object still dominates the radiation field in
its immediate vicinity. From the total intensity images, we find that Mira~A is brighter than Mira~B by factor of $\sim 45$
and $\sim 1200$ at $0.65~\mu$m and $0.82~\mu$m, respectively.
By considering the separation between the two stars of $\sim 90$~au \citep[][for an assumed distance of 107~pc]{Ireland2007},
we estimate that Mira~B dominates the radiation field
at $0.65~\mu$m and $0.82~\mu$m up to distances of $\sim 13$~au and $\sim 2.5$~au in the direction of Mira A, respectively.
This is reflected in the orientation of the polarization vectors around Mira~B, which are mostly tangential to a circle centred on Mira~B at 0.65~$\mu$m and to a circle
centred on Mira~A at 0.82~$\mu$m (see Fig.~\ref{fig:polVectors}). The polarization vectors appear tangentially aligned to Mira~B{ at 0.65~$\mu$m}
over a region of $\sim 60$~mas in radius
around the companion star. This corresponds to ${\sim 6}$~au and is smaller by a factor of two than the estimate above.
We note that determining the direction of the polarization vectors in that region is not straightforward.
For instance, the polarization vectors to the west of Mira~B in the image in filter NR
are radially aligned with respect to Mira~B, which we attribute
to the beam-shift effect (see Sect.~\ref{sec:beam-shift}).

\subsection{Gas model}
\label{sec:columnDens}

 \begin{figure*}[t]
   \centering
      \includegraphics[width= 17cm]{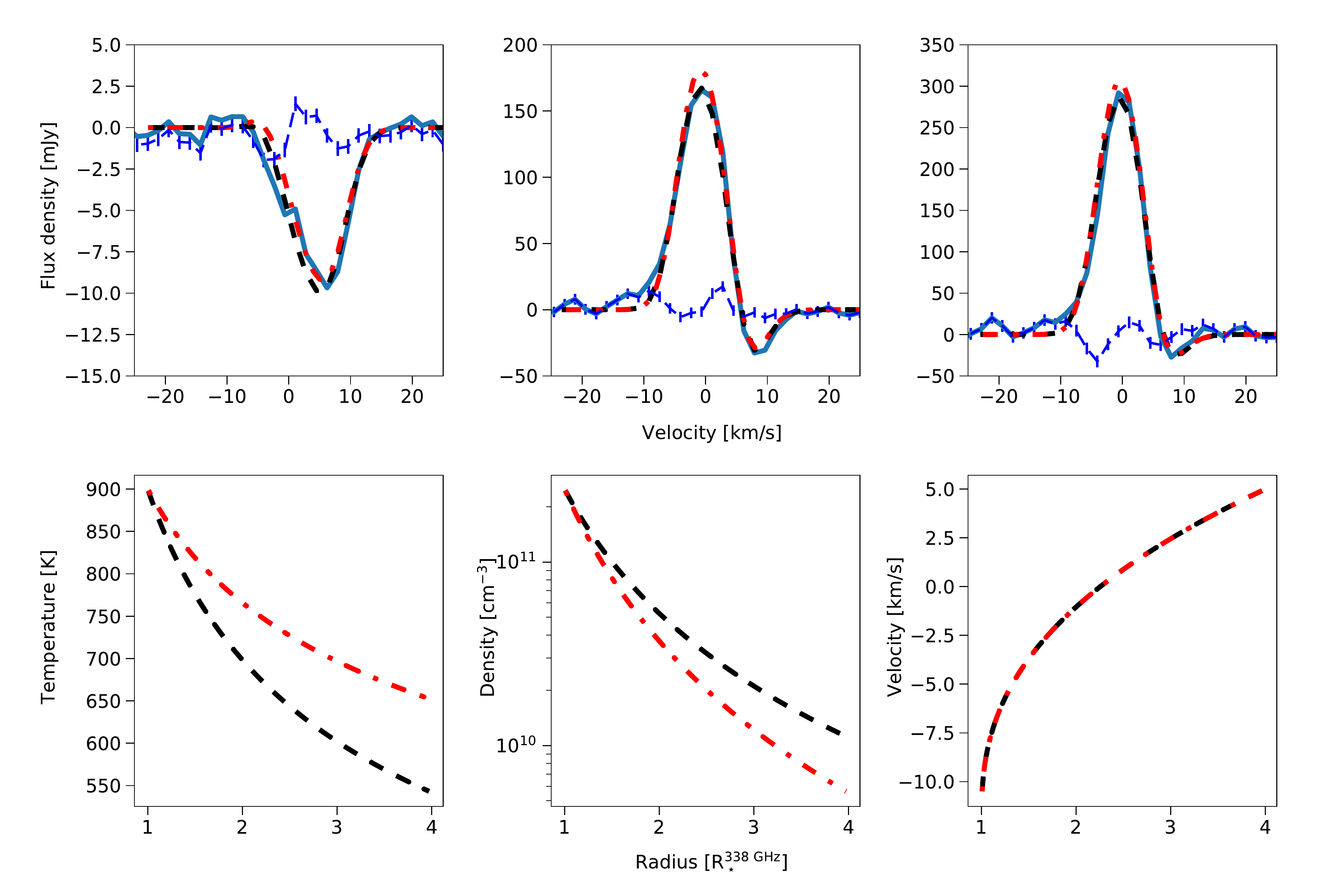}
      \caption{{ Best model fit to the CO~$v=1, J=3-2$ line observed using ALMA (dashed black line) and preferred model after constraints from the $^{13}$CO~$J=3-2$ line are
      considered (dot-dashed red line). The parameters of these two models are given in Table~\ref{tab:vibCO-model}.
       Upper panels: Observed spectra of the CO~$v=1, J=3-2$ line (solid blue line) and residuals (thin blue line with error bars) obtained by subtracting the best-fit model
       from the observed spectra. The spectra were extracted from
      circular regions with radii of $20$~mas (left), $50$~mas (middle), and $100$~mas (right). Lower panels: Gas temperature (left),
      density (middle), and velocity profiles (right) of the two models. The velocity is positive for material flowing away from the stellar surface.}}
         \label{fig:vibCO-model} 
   \end{figure*}

Emission from the $N=9-8$ transition of AlO and from several transitions of TiO$_2$ (see Table~\ref{tab:obs}) is detected in the ALMA observations. These molecules contain elements
thought to be important for dust formation (aluminium and titanium) and an analysis of the observed emission in comparison to the polarized light is particularly
interesting. Doing a similar analysis for Si from the data at hand is not
straightforward, because emission from the $^{29}$SiO $J=8-7$ line is very optically thick, and the emission becomes
insensitive to column density in this regime. Observations of lower abundance isotopologues of SiO might be the best way to constrain the SiO abundance distribution. 
To estimate the total column density of gas, we use the lines $v=1, J=3-2$ of $^{12}$CO and $v=0, J=3-2$ of $^{13}$CO.
We originally intended to also use the $v=0, J=3-2$ line of C$^{18}$O, but our data show that
it seems to be weakly masing, { as inferred from} the spatially compact and spectrally narrow emission observed.

\subsubsection{CO~$v=1, J=3-2$}
\label{sec:modelCO}

We modelled the vibrationally-excited CO $J=3-2$ line with the same code and approach as used by \cite{Khouri2016b} and \cite{Vlemmings2017}.
In the code, local thermodynamical equilibrium is assumed to calculate the level populations{ in a shell around the} central star.{ We assume a CO abundance of $4 \times 10^{-4}$ relative to H$_2$.}
The velocity, density, and temperature of the gas are defined by power laws and spherical symmetry is assumed. After the level populations
are obtained, images at different velocities are calculated by solving the radiative transfer equation at several lines of sight spread radially
from the centre of the star to the edge of the{ molecular shell}. We convolved the model images with the beam of the ALMA observations
obtained by us and extracted model spectra from different apertures to compare with the observations.

The model star was assumed to be a uniform disc when projected on the plane of the sky. The stellar radius and temperature were set to $R^{\rm 338 GHz}_\star = 2.15$~au
and $T_\star = 2100$~K based on our fit to the continuum visibilities. We adopt a temperature profile of the type
\begin{equation*}
T_{\rm kin}(r) = T_\circ (R^{\rm 338~GHz}_\star/ r)^\epsilon .
\end{equation*}
An initial input for the values of $T_\circ$ and $\epsilon$ was obtained using the brightness temperature of the $^{13}$CO line at the $\upsilon_{\rm LSR}$ of the system.
We note that the observed brightness temperature might be affected by resolved-out flux.
The flux loss is larger for the intensity at $\upsilon_{\rm LSR}$ than for the integrated flux{ (discussed in Sect.~\ref{sec:MaxScale})},
because at the $\upsilon_{\rm LSR}$ the emission region is larger.
We estimate the resolved-out flux in the $^{13}$CO line using the observations of \cite{Ramstedt2014}
and find that 10\% of the flux density per beam,
at $\upsilon_{\rm LSR}$, from the compact molecular line emission region is not recovered by our observations. 
Therefore, we calculated brightness temperature maps correcting for this flux loss, and we find the brightness temperature
profile to be described by $T_\circ = 900$~K and $\epsilon = 0.4$. 
The outer radius of the modelled region (of 4~$R^{\rm 338~GHz}_\star$, or 84~mas)
was defined based on the observed emission region of the CO~$v=1, J=3-2$ line.

The velocity and density profiles are defined by
\begin{equation*}
\upsilon(r) = \upsilon_\circ+(\upsilon_{\rm f}-\upsilon_\circ) [(r-R^{\rm 338~GHz}_\star)/(R_{\rm out}-R^{\rm 338~GHz}_\star)]^\beta
\end{equation*}
and
\begin{equation*}
n(r) = n_\circ \times (R^{\rm 338~GHz}_\star/r)^\eta,
\end{equation*} respectively.
The free parameters in the model that were varied to fit the data are the stochastic velocity, $\upsilon_{\rm sto}$, the velocity at $R^{\rm 338~GHz}_\star$,
$\upsilon_\circ$, the velocity at $R_{\rm out}$, $\upsilon_{\rm f}$, the exponent of the velocity law, $\beta$, the density at $R^{\rm 338~GHz}_\star$, $n_\circ$, the exponent of the
density law, $\eta$, the temperature at $R^{\rm 338~GHz}_\star$, $T_\circ$, and the exponent of the temperature law, $\epsilon$.
The velocity of Mira with respect to the local standard of rest, $\upsilon_{\rm LSR}$, was also
adjusted to fit the observations. We find $\upsilon_{\rm LSR} = 47.7\pm0.1$~km/s.

{ Our fitting procedure was carried out as follows.
We first searched for the region of the parameter space that provided the best fits to the data by eye. Then, we
calculated grids of models to refine the best models and to estimate uncertainties on the obtained parameters.
The best models were chosen by minimizing the $\chi^2 \equiv \sum_{i=1}^{\rm n} (S_i^{\rm obs} - S_i^{\rm mod})^2/\sigma^2_i$, where $S_i^{\rm obs}$
and $S_i^{\rm mod}$ are the observed and modelled flux density values and $\sigma_i$ is the standard deviation within a given observed spectrum.  
In the grid calculations, we varied all the free parameters, $T_\circ$, $\epsilon$, $\upsilon_{\rm sto}$, $\upsilon_\circ$, $\upsilon_{\rm f}$, $\beta$,
$n_\circ$, $\eta$, and $\upsilon_{\rm LSR}$. The grid was progressively refined until uncertainties on the parameters were well defined.
We calculated close to $2\times10^4$ models spanning the ranges given in Table~\ref{tab:vibCO-model}.
The fits had 54 degrees of freedom (DF), which is the number of fitted spectral channels minus the number of free parameters.
Our best fit reached a reduced-$\chi^2 = \chi^2/{\rm DF} \sim 4.0$. This shows that a more complex model (non-spherically symmetric) is required to
fully reproduce the exquisite ALMA images of the molecular gas emission close to Mira.
Nonetheless, we are able to constrain the free parameters in the context of our simplified model.}

The best fitting model found by us is shown in Fig.~\ref{fig:vibCO-model} and the corresponding
parameters{ and uncertainties} are given in Table~\ref{tab:vibCO-model}. In our best model, the gas density decreases{ from $2.5^{+0.8}_{-0.5} \times 10^{11}~{\rm cm}^{-3}$
at $r=R^{\rm 338~GHz}_\star$~(21~mas) to $(1.0 \pm 0.5) \times 10^{10}~{\rm cm}^{-3}$
at $r = 4~R^{\rm 338~GHz}_\star$ (84~mas).} The model requires infalling material close to the star to reproduce the observed red-shifted absorption (see Fig.~\ref{fig:vibCO-model}).
{ As expected, the free parameters are not all independent. For instance, lower values of $T_\circ$ require lower values
of $\epsilon$ and vice-versa. The same is true for $n_\circ$ and $\eta$. We find a mass in the vibrationally-excited CO region of $(3.8 \pm 1.3) \times 10^{-4} M_\odot$.}

The large step between the continuum brightness temperature and the gas kinetic temperature at the stellar radius is a somewhat inconsistent feature
of our model. We explored making the gas kinetic temperature at $R^{\rm 338~GHz}_\star$ (2.15~au) equal to the temperature of the model star (2100~K)
while increasing the value of the exponent of the temperature power law, so that the temperature would drop to ${\sim 500}$~K
at ${r \sim 4~R^{\rm 338~GHz}_\star}$. We find that models with such a temperature profile are unable to simultaneously reproduce the absorption towards
the star and the emission from around it, because the absorption becomes too weak, in a relative sense. This shows that our model requires
fairly cold gas, $T \sim 1000$~K,
already very close to the star. One way to make our model more consistent would be to introduce a break in the temperature power law, so that the
temperature profile would be steep very close to the star and more shallow after reaching $T \sim 1000$~K. We have not explored this possibility, however,
to avoid increasing the number of free parameters in the model.

Our model assumes
a sharp transition between the region where
continuum opacity dominates and the region where CO~$v=1, J=3-2$ line opacity dominates, while in reality the transition could be gradual.
We note, however, that we obtain a good fit to the stellar disc of Mira at 338~GHz using a uniform disc, and, therefore, there is no
evidence for a slow radial decrease of continuum opacity at the time of the observations.
Hence the large difference between the stellar brightness temperature and the gas kinetic temperature at $R^{\rm 338~GHz}_\star$
indicates a sharp decline of the kinetic temperature from $\sim 2100$~K to $\sim 1000$~K at the radius of the 338~GHz continuum, $R^{\rm 338~GHz}_\star$.
This result seems in qualitative agreement with the hydrodynamical models of
\cite{Ireland2011}, since at some phases a sharp temperature decline of approximately the required magnitude is observed at $\sim 2$~au.

The total mass we find in the CO~$v=1, J=3-2$ line emission region, $(3.8 \pm 1.3) \times 10^{-4} M_\odot$, is almost two orders
of magnitude larger than that found by 
\cite{Khouri2016b}, $\sim 6\times10^{-6}~M_\odot$,
by modelling spatially-unresolved observations of the CO~$v=1, J=3-2$ line towards Mira at roughly the same pulsation phase and using a similar approach
to the one in this work. Although variability might explain, at least partially, the different masses found, the lack of spatial information in 
their study also plays an important role.
Since the star was unresolved in the observations used by the above authors, they assumed a stellar brightness temperature of 3000~K
to derive the stellar radius at sub-millimetre wavelengths. This stellar brightness temperature
is clearly too high based on the observations we report. This leads to an underestimation of the stellar radius by a factor of {\bf approximately two}.
The model by these authors was also simpler than ours, since the authors assumed constant excitation temperature and gas density over the line-formation region.
The values of the excitation temperature they considered range from 1000~K to 2000~K, which are higher than those we find.
This shows how spatial information
provides crucial constraints for this type of study.

The gas density we find at $r=21$~mas, $2.5^{+0.8}_{-0.5} \times 10^{11}~{\rm cm}^{-3}$, is almost two orders of magnitude
lower than that obtained by \cite{Wong2016}, $10^{13}$ cm$^{-3}$, at the same region {\bf by calculating models to fit ALMA observations of H$_2$O and SiO
line emission acquired during science verification
time in the 2014 ALMA Long Baseline Campaign.}
At $r=84$~mas our result, $(1.0 \pm 0.5) \times 10^{10}~{\rm cm}^{-3}$, is more comparable to theirs, $\sim 5 \times 10^{10}~{\rm cm}^{-3}$.
This could be because of variability, since gas with radial speeds of ${\sim 8}$~km/s would be able to travel through this region
in the period of three years that separates the observations we report from those of \cite{Wong2016}.

The hydrodynamical models calculated by \cite{Ireland2011} for the parameters of Mira produce gas densities that are very variable
through the different pulsation cycles. The values presented vary between approximately
$10^{12}~{\rm cm}^{-3}$ and $2.5 \times 10^{10}~{\rm cm}^{-3}$
at 2.15~au and $10^{9}~{\rm cm}^{-3}$ and $3 \times 10^7~{\rm cm}^{-3}$ at 5~au (which is the maximum radius for which gas densities are presented).
While the density we find at 2.15~au ($r=R^{\rm 338~GHz}_\star$), $(2.5^{+0.8}_{-0.5}) \times 10^{11}~{\rm cm}^{-3}$ is in agreement with the values reported by \cite{Ireland2011},
we find higher densities at 5~au, $(3.0 \pm 1.0) \times10^{10}~{\rm cm}^{-3}$, than those their models predict.

 \begin{figure}[t]
   \centering
      \includegraphics[width= 6.5cm]{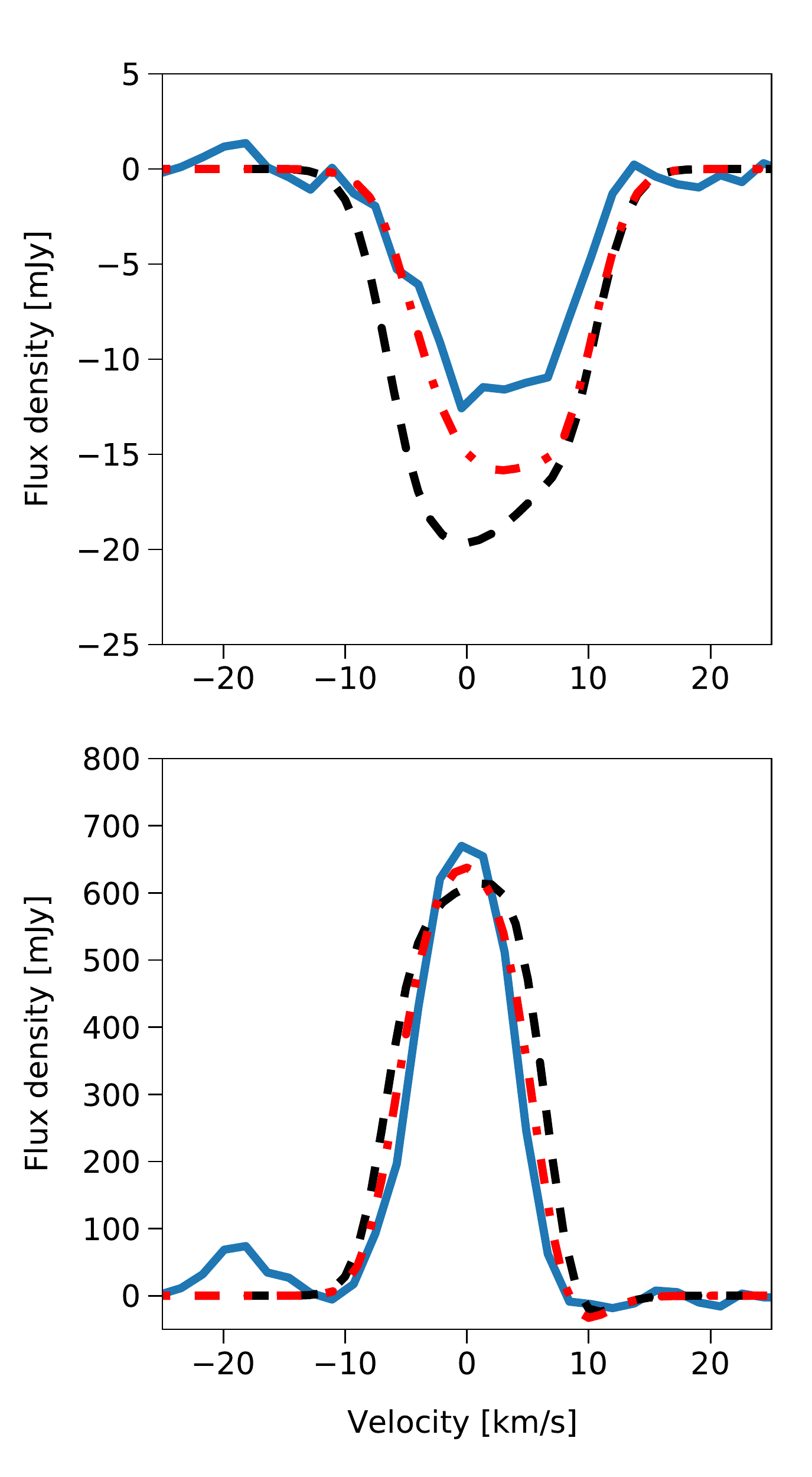}
      \caption{ Comparison between the observed $^{13}$CO~$J=3-2$ line and
      predictions from the models that provide good fits to the CO~$v=1, J=3-2$ line. The dashed black line shows the best-fitting model to only the CO~$v=1, J=3-2$ line,
      and the dot-dashed red line shows the preferred model after the analysis of the $^{13}$CO~$J=3-2$ line.
      The models are the same as shown in Fig.~\ref{fig:vibCO-model} and with the parameters given in Table~\ref{tab:vibCO-model}.
      The upper and lower panels show spectra extracted using apertures with radii equal to 20 and 84~mas, respectively.}
         \label{fig:13CO-model} 
   \end{figure}

\subsubsection{Further constraints with the $^{13}$CO, $J=3-2$ line}

{ Now we compare models to the observed $^{13}$CO~$v=0, J=3-2$ line based on the results from the CO~$v=1, J=3-2$ line calculations.
The $^{13}$CO~$v=0, J=3-2$ models were computed
assuming local thermodynamical equilibrium (LTE), as for the CO~$v=1, J=3-2$ line.
We extracted the $^{13}$CO~$v=0, J=3-2$ line spectrum using circular apertures with
radii of 20 and 84~mas ($4~R^{\rm 338~GHz}_\star$).
The area of the vibrationally-excited
CO region corresponds to ${\sim 40\%}$ of the total area of the compact molecular line emission region, while this region accounts for
approximately $50\%$ of the flux we recover in the $^{13}$CO line.

We calculated models considering the best-fit models, within the uncertainties given in Table~\ref{tab:vibCO-model}, and assuming a 
$^{12}$C-to-$^{13}$C ratio of $10 \pm 3$ \citep{Hinkle2016}. We find that we obtain reasonably good fits when using models with masses at
the lower end of the one-$\sigma$ uncertainty interval we obtain from the CO $v=1, J=3-2$ line,
of $(3.8 \pm 1.3) \times 10^{-4}~M_\odot$. These models have steeper density profiles and shallower temperature
profiles than the best-fitting model from our grid but, nonetheless, still provide a reasonable fit to the CO $v=1, J=3-2$ observations (see Fig.~\ref{fig:vibCO-model}).
Our fit improves by considering a $^{12}$C-to-$^{13}$C ratio of 13,
at the edge of the one-$\sigma$ uncertainty interval given by \cite{Hinkle2016}. The preferred model fit, taking into account the $^{13}$CO line, is
compared to the data in Fig.~\ref{fig:13CO-model} and its parameters are given
in Table~\ref{tab:vibCO-model}. Based on the model predictions for the $^{13}$CO~$J=3-2$ line, we conclude that the mass in the
vibrationally-excited CO region is $\sim 2.5 \times 10^{-4}~M_\odot$. This corresponds to a total H$_2$
column density of $3.4 \times 10^{24}$~cm$^{-2}$ in this inner region.}

 \begin{figure}[t]
   \centering
      \includegraphics[width= 8.5cm]{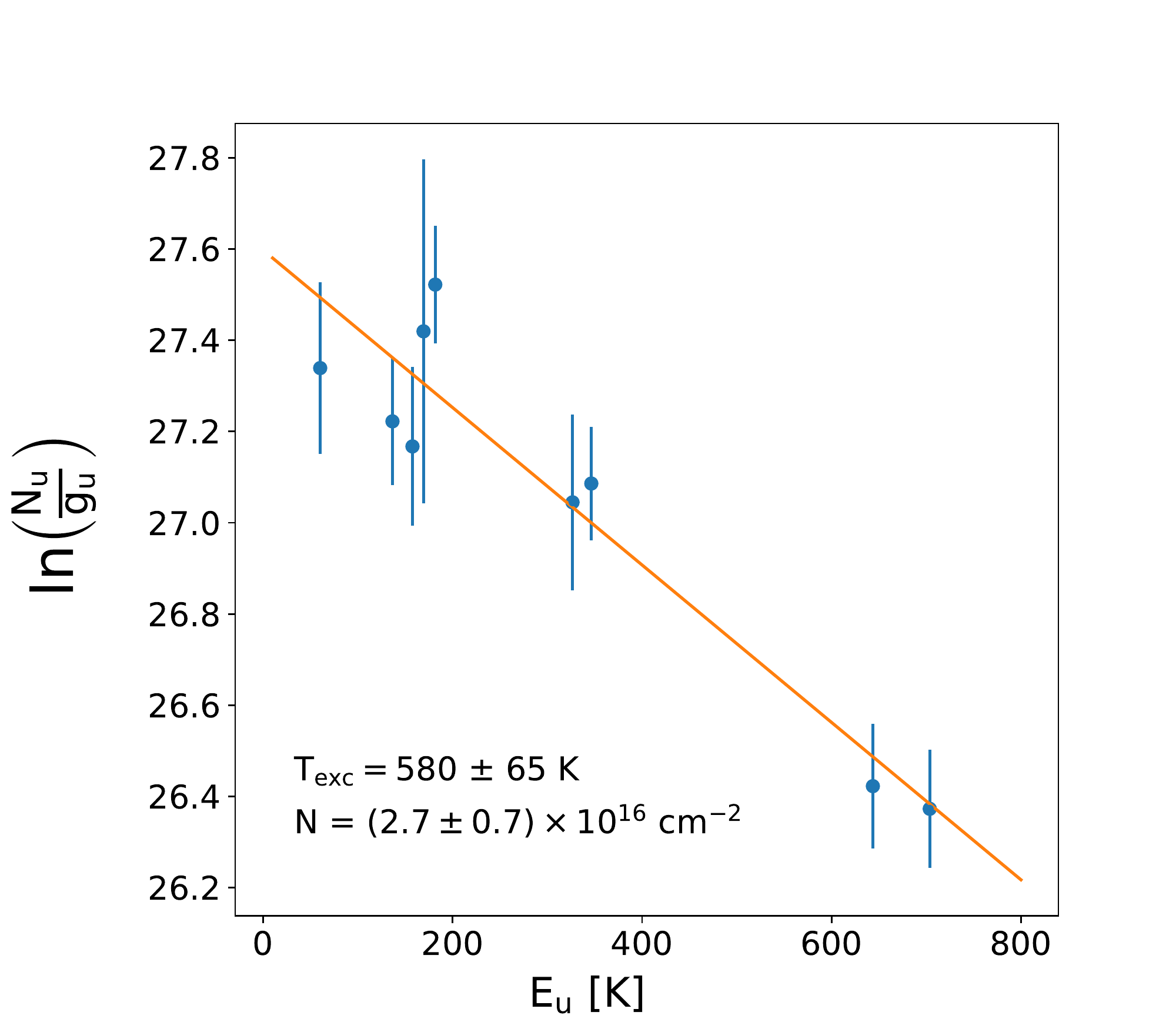}
      \caption{Rotational diagram obtained using the unblended TiO$_2$ lines given in Table~\ref{tab:obs}.}
         \label{fig:TiO2_rotDiag}
   \end{figure}

\subsubsection{Depletion of dust-forming elements}

{ Having constrained the total amount of gas, we are now able to study the abundances of molecules that bear dust-forming elements,
TiO$_2$ and AlO.}
Of the 26 lines of TiO$_2$ that fall within the frequency range of the ALMA observations \citep[based on measurements by ][]{Kania2011},
about 20 can be identified in the spectrum. Of these, most are blended
with stronger or equally strong lines. We identified nine lines for which the integrated line flux could be determined reliably.
{ We extracted the fluxes using a circular aperture with $4~R^{338~{\rm GHz}}_\star$ radius (see Table~\ref{tab:obs}).
For TiO$_2$ lines with $E_{\rm u} > 160~K$, the vibrationally-excited CO region accounts for $\gtrsim 70\%$ of the total flux from the compact molecular line
emission region, while for the three lines with the lowest excitation energy this fraction is between 50 and 60\%.
From the rotational diagram shown in Fig.~\ref{fig:TiO2_rotDiag},
we derive an excitation temperature of $580 \pm 65$~K and a column density of $(2.7 \pm 0.7) \times 10^{16}$~cm$^{-2}$.
We note that the excitation temperature obtained is comparable to the lowest values obtained in the CO model at $r \sim 4~R^{\rm 338~GHz}_\star$.
The column density we find is a factor of 2.5 larger than the value of ${N_{\rm TiO_2} = (1.15\pm0.12)\times10^{16}}$~cm$^{-2}$ obtained by \cite{Kaminski2017}
using ALMA observations acquired in 2015. These authors also found a significantly lower excitation temperature of $174\pm7$~K.
Their results are based on a region similar to the vibrationally-excited CO region we consider.
This difference could be explained by variability of the source, or by uncertainties and differences between the two analyses.
Interestingly, for TiO, \citeauthor{Kaminski2017} find an excitation temperature of $491\pm96~K$, which is comparable to what we find for TiO$_2$.
Using the H$_2$ column density obtained from the CO model, our results imply a TiO$_2$-to-H$_2$ ratio of $\sim 8 \times 10^{-9}$.

Emission from the AlO~$N=9-8$ line peaks very close to the star. As shown in Fig.~\ref{fig:vibCO+AlO},
roughly 60\% of the emission of the compact
molecular line emission region arises from the vibrationally-excited CO region.
Assuming LTE excitation and an excitation temperature of $580\pm65$~K, as found for TiO$_2$,
we find that an AlO column density of $(7 \pm 2) \times 10^{15}~{\rm cm}^{-2}$ is needed to reproduce the
observed emission from the vibrationally-excited CO region ($r \leq 4~R^{\rm 338 GHz}_\star$).
This translates to an AlO-to-H$_2$ ratio of $\sim 2 \times 10^{-9}$.
Assuming a higher excitation temperature of 1000~K would cause the column density and AlO-to-H$_2$ ratio to increase only by a factor of two.
The column density we find is comparable to that reported by \cite{Kaminski2016}, $\sim 5 \times 10^{15} $~cm$^{-2}$, for an excitation temperature
of $329 \pm 51$~K.

 \begin{figure}[t]
   \centering
      \includegraphics[width= 8.6cm]{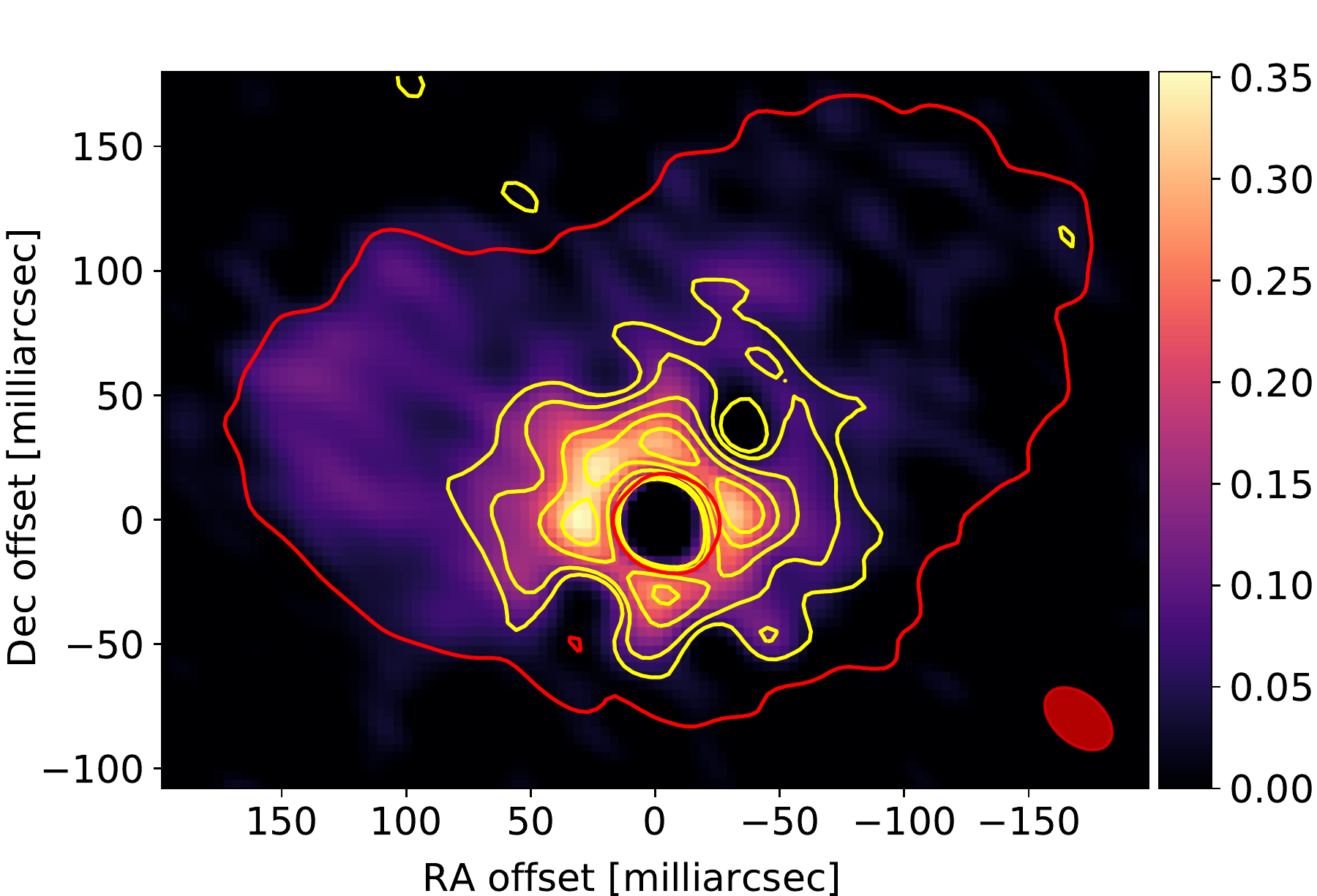}
      \caption{Comparison between the AlO~$N=9-8$ and the CO~$v=1, J=3-2$ lines observed using ALMA.
      The colour map shows the AlO~$N=9-8$ line in ${\rm Jy~km~s^{-1}~beam^{-1}}$. The yellow contours show
      emission in the CO~$v=1, J=3-2$ line at 80\%, 60\%, 40\%, 20\%, and 10\% levels of the peak emission.
      The red contour shows the 10\% level of the SO emission (same
      as in Fig.~\ref{fig:polDeg+SO}).
      { The offsets in declination and right ascension are given with respect to the position of Mira~A.}}
         \label{fig:vibCO+AlO}
   \end{figure}

The Ti and Al abundances with respect to H$_2$ for a gas with solar composition \citep{Asplund2009} and in which hydrogen is fully molecular are $1.8 \times 10^{-7}$
and $5.6 \times 10^{-6}$, respectively.
Therefore, the TiO$_2$ molecules account for $\sim 4.5\%$ of the expected titanium and the AlO molecules account for $< 0.1\%$ of the expected aluminium.
\cite{Kaminski2017} found a 3.4 times higher column density of TiO than of TiO$_2$ towards Mira.
If we assume the same ratio is valid at the epoch of our observations, our results indicate that these two molecules could account for close to a fifth of
all titanium. Hence our results indicate that a significant fraction of titanium is locked in other carriers. We note that the lack of TiO lines in our data and the relatively large
uncertainties make a firm conclusion on the combined abundances of TiO and TiO$_2$ not possible, however. Moreover, observations by
\cite{Kaminski2017}
revealed that atomic Ti also exists in the extended atmosphere of Mira~A, but the authors were unable to derive the Ti column density.
Hence we conclude that the titanium not found in TiO and TiO$_2$ molecules may be in atomic form or in dust grains. Determining the amount of free Ti atoms is therefore
necessary for measuring the depletion of this element.
The low abundance of AlO makes a scenario possible in which aluminium is very efficiently depleted into dust, but the uncertainties
on the abundance of other aluminium-bearing
species also hampers a firm conclusion.
Nonetheless, thanks to the relatively larger solar abundance of aluminium,
we are able to constrain the amount of Al atoms that can be locked in grains of different sizes by studying
the polarized light observations with ZIMPOL (see below).}
An alternative explanation for the low estimated column density of AlO could be that the assumption of LTE is not valid.
We consider this to be unlikely, however, because the high gas densities we derive in the region ($> 10^9$~cm$^{-3}$)
imply that the level populations of AlO are thermalized \citep{Kaminski2016}.
 \begin{table*}
\footnotesize
\setlength{\tabcolsep}{12pt}
\caption{TiO$_2$ transitions that fall within the spectral range covered by the ALMA observations reported by us. We give the integrated flux
of the detected lines obtained by integrating emission over a circular aperture with 84~mas radius (see text). The blended lines are indicated, and
when the blending line is not known, we refer to it as UD (unidentified line).}              
\label{tab:obs}      
\centering                                      
\begin{tabular}{c c c c c c c}
Transition & $\nu$ & $E_{\rm u}$ & Flux & $\sigma_{\rm Flux}$ & $g_{\rm u}$ & $A_{\rm u,l}$ \\
& [GHz] & [K] & [$10^{-20}$ W/m$^2$] & [$10^{-20}$ W/m$^2$] & & [s$^{-1}$] \\
\hline
31(4,28) -- 31(3,29) & 329.598  & 376.53 & \multicolumn{2}{c}{\it Blended with UD} & 63 & $2.85\times 10^{-3}$ \\ 
50(8,42) -- 50(7,43) & 329.704 & 1023.25 & \multicolumn{2}{c}{\it Not detected} & 101 & $5.13 \times 10^{-3}$ \\ 
35(20,16) -- 36(19,17) & 330.359 & 894.53 & \multicolumn{2}{c}{\it Not detected} & 71 & $4.63 \times 10^{-3}$\\%

27(8,20) -- 27(7,21) & 331.212 & 346.43 & $1.20$ & $0.15$ & 55 & $4.56\times 10^{-3}$\\
26(8,18) -- 26(7,19) & 331.240 & 326.60 & $1.10$ & $0.23$ & 53 & $4.52\times 10^{-3}$ \\
39(8,32) -- 39(7,33)$^{\rm a}$ & 331.382 & 643.64 & $0.98$ & $0.14$ & 79 & $5.03\times 10^{-3}$\\

\phantom{0}12(3,9) -- 11(2,10) & 331.600 & 67.55 & \multicolumn{2}{c}{\it Blended with UD} & 25 & $1.35\times 10^{-3}$\\
14(4,10) -- 14(1,13) & 332.355 & 94.54 & \multicolumn{2}{c}{\it Not detected} & 29 & $1.77\times 10^{-4}$\\
32(19,13) -- 33(18,16) & 332.528 & 778.07 & \multicolumn{2}{c}{\it Not detected} & 65 & $4.35\times 10^{-4}$\\

17(8,10) -- 17(7,11) & 341.463 & 181.68 & $1.10$ & $0.15$ & 35 & $4.12\times 10^{-3}$ \\

32(3,29) -- 32(2,30)$^{\rm b}$ & 341.594 & 399.00 & \multicolumn{2}{c}{\it Not detected?} & 65 & $3.08\times 10^{-3}$\\

16(8,8) -- 16(7,9)      & 341.876 & 169.29 & $0.91$ & $0.40$ & 33 & $4.00 \times 10^{-3}$\\
41(8,34) -- 41(7,35)$^{\rm d}$ & 342.120 & 703.68 & $1.08$ & $0.15$ & 83 & $5.47\times 10^{-3}$\\
15(8,8) -- 15(7,9)$^{\rm e}$ & 342.217 & 157.63 & $0.64$ & $0.12$ & 31 & $3.85 \times 10^{-3}$ \\
9(5,5) -- 8(4,4) & 342.345 & 60.23 & $0.57$ & $0.10$ & 19  & $4.71\times 10^{-3}$\\

14(8,6) -- 14(7,7) & 342.491 & 146.70 & \multicolumn{2}{c}{\it Blended with SiO $v=2, J=8-7$} & 29 & $3.67\times 10^{-3}$ \\

13(8,6) -- 13(7,7) & 342.708 &136.51 & $0.53$ & $0.08$ & 27 & $3.46 \times 10^{-3}$ \\

6(6,0) -- 5(5,1) & 342.862 & 54.87 & \multicolumn{2}{c}{\it Blended with UD} & 13 & $7.92\times 10^{-3}$\\
12(8,4) -- 12(7,5) & 342.877 & 127.05 & \multicolumn{2}{c}{\it Blended with UD} & 25 & $8.43\times 10^{-3}$\\
11(8,4) -- 11(7,5) & 343.004 & 118.33 & \multicolumn{2}{c}{\it Blended with $^{29}$SiO $v=0, J=8-7$} & 23 & $2.86\times 10^{-3}$\\
46(10,36) -- 46(9,37)\phantom{0} & 343.049 & 903.83 & \multicolumn{2}{c}{\it Not detected}& 93 & $5.97\times 10^{-3}$\\
23(3,21) -- 22(2,20) & 343.071 & 208.30 & \multicolumn{2}{c}{\it Blended with $^{33}$SO $N_J = 9_8 - 8_7$ and UD} & 47 & $5.27\times 10^{-3}$\\
10(8,2) -- 10(7,3) & 343.097 & 110.33 & \multicolumn{2}{c}{\it Blended with $^{33}$SO $N_J = 9_8 - 8_7$} & 21 & $2.43\times 10^{-3}$ \\
9(8,2) -- 9(7,3) & 343.162 & 103.06 & \multicolumn{2}{c}{\it Blended with UD} & 19 & $1.86\times 10^{-3}$\\
8(8,0) -- 8(7,1) & 343.204 & 96.52 & \multicolumn{2}{c}{\it Blended with UD} & 17 & $1.09\times 10^{-3}$\\
13(4,10) -- 12(3,9)\phantom{0} & 343.243 & 84.03 & \multicolumn{2}{c}{\it Blended with UD} & 27 & $2.97\times 10^{-3}$\\
\end{tabular}
\tablefoot{a - The expected transition at 331.382~GHz was matched to an observed line centred at 331.386~GHz. Since the error on the tabulated frequency was relatively large (2.4~MHz),
we identify the line at 331.386~GHz with transition 39(8,32) -- 39(7,33). b - A line with flux that fits in the observed rotational diagram was observed at
341.600~GHz. Since the tabulated uncertainty on the transition frequency is relatively small (0.9~MHz), we have not included the line in the fitting procedure.
d - The expected transition at 342.120~GHz was matched to an observed line centred at 342.129~GHz. Since the error on the tabulated frequency was relatively large
(2.6~MHz), we identify the line at 342.120~GHz with transition 41(8,34) -- 41(7,35). e - Blended with a significantly weaker line.}
\end{table*}

\subsubsection{The aluminium budget}

\cite{Gobrecht2016} calculated the abundances of aluminium-bearing species
and of aluminium oxide dust particles considering a shock-induced chemistry in the extended atmosphere of a model AGB star.
The model was tailored to fit the star IK~Tau, with mass-loss rate and gas densities in the inner wind at least an order of magnitude larger than those of Mira.
They found that a much larger fraction of the aluminium can be in AlOH than in AlO, and that AlOH can account for $\sim 30\%$ of all aluminium
at a few stellar radii. Observations of Mira have suggested that AlOH is located only very
close to the stellar photosphere, however, at temperatures of $\sim 2000$~K \citep{Kaminski2016}. Although
the authors were unable to constrain the column density of AlOH, the abundance does not seem to be as high as expected based on the models of \cite{Gobrecht2016}.
\cite{Decin2017} observed two AGB stars, one with a higher mass-loss rate than Mira \citep[and the same one modelled by][IK~Tau]{Gobrecht2016}
and another with a mass-loss rate more comparable to that of Mira (R~Dor).
They found that the abundances of AlO and AlOH are much lower than the expected aluminium abundance and that these molecules together
account for only 2\% of the total aluminium.
This does not support the high abundance of AlOH predicted by \cite{Gobrecht2016}. Moreover, the abundance of AlO in R~Dor was found to be almost two
orders of magnitude larger than that of AlOH.

The emission from AlOH gas in the vibrationally-excited CO region can be estimated based on our gas model.
Assuming solar composition, we find that even if only $\sim 10$\% of aluminium is in AlOH molecules (or an AlOH abundance
$\sim 2.8 \times 10^{-7}$), AlOH emission
would produce a strong, optically-thick $N=11-10$ line at 346.155~GHz with a flux density $\sim 1$~Jy. The signal estimated for only this small region
is about an order of magnitude larger than reported by \cite{Kaminski2016} within the beam of the Atacama Pathfinder Experiment (APEX)
telescope. Therefore, we conclude that AlOH most likely does not
account for a significant fraction of aluminium atoms in the close environment of Mira~A.
The aluminium budget is, therefore, still very incomplete. As discussed by \cite{Kaminski2016}, aluminium
is expected to be in atomic form at temperatures $\gtrsim 2200$~K, { which is considerably higher than the gas temperatures we derive}. In broad terms,
the search for other aluminium-bearing molecules has not revealed clear candidates for the bulk of aluminium \citep{Kaminski2016,Decin2017}.

\subsection{Aluminium oxide dust grains and molecular clusters}
\label{sec:modelDust}

A very likely explanation of the low column density of AlO is that aluminium atoms are very efficiently depleted into dust.
Mid-infrared interferometric observations of dust thermal emission towards other Mira variables
show aluminium-bearing dust to form at radii $\lesssim 2$~R$_\star$ \citep[e.g. ][]{Zhao-Geisler2012,Karovicova2013},
models for dust emission suggest aluminium oxide grains must exist in the extended
atmosphere \citep{Khouri2015},
and dust condensation models predict aluminium oxide dust to form efficiently from the gas phase \citep{Gobrecht2016,Hoefner2016}.
Therefore, in this section we study the polarized light observed using ZIMPOL in order to constrain the presence and properties of aluminium oxide dust grains
that are expected
to exist close to the star.

{ Theoretical models for dust nucleation and growth find that aluminium oxide grains can form efficiently around O-rich AGB stars.}
\cite{Gobrecht2016} predict a significant fraction of Al atoms ($\sim 30\%$) to be found in aluminium oxide dust grains (Al$_2$O$_3$),
provided that the particles can withstand the passage of shocks. Moreover, they found that such grains grow to sizes between 0.1 and 0.3~$\mu$m
within 2~$R_\star$, with $R_\star = 1.7$~au in their models.
\cite{Hoefner2016} calculated the growth of amorphous aluminium oxide dust on top of seed dust particles considering a small range of input parameters
of model O-rich stars that produce mass-loss rates of the same order as that of Mira. They found that
more than 50\% of the aluminium condenses
into dust grains{ , if the grains have low absorption opacity. The particles reach
sizes} $\gtrsim 0.1~\mu$m at ${\sim 1.7~R_\star}$, with the stellar radius in their models being equal to either 1.4 or 1.8~au for different models.
To constrain the size of possible
aluminium oxide dust particles located in the vibrationally-excited CO region,
we calculated continuum radiative-transfer models using MCMax \citep{Min2009}
in its spherically-symmetric mode and with a similar approach to that of \cite{Khouri2016}.

\subsubsection{The lattice structure and optical constants of Al$_2$O$_3$}

The optical constants of the considered solid have a major effect on the results of the calculations, because they directly determine the scattering and absorption
properties of the model grains. We discuss the optical properties of Al$_2$O$_3$ solids based on the complex index of refraction,
$m(\lambda) = n(\lambda) + i \kappa(\lambda)$. The real and imaginary parts of the index of refraction mostly affect the scattering and absorption properties
of the grains, respectively.
In this work, we are mainly interested in the value of
the real part of the index of refraction, $n$, at 0.82~$\mu$m.

 \begin{table}
\footnotesize
\setlength{\tabcolsep}{12pt}
\caption{Summary of optical constants of aluminium oxide solids in visible and NIR wavelengths.}
\label{tab:OptConst}
\centering
\begin{tabular}{l@{ \phantom{ }} c l l}
\hline
\hline
Reference & Struct. & $n_{0.82~\mu{\rm m}}$ & $\kappa_{\rm V+NIR}$ \\
\hline
\citeauthor{Suh2016} & A & ${\sim 1.45}$ & $\sim 0.2$ \\
\citeauthor{Harman1994} & C ($\alpha$) & $\sim 1.7$ & $< 0.1$\\
\citeauthor{Koike1995} (Comb.) & C ($\gamma$) & 1.59 & $1$ -- $4 \times 10^{-2}$ \\
\citeauthor{Koike1995} (Comm.) & C ($\gamma$) & 1.56 & $0.5$ -- $1 \times 10^{-2}$ \\
\citeauthor{Edlou1993} (Sputt.) & ? & 1.64 & $< 4 \times 10^{-4}$ \\
\end{tabular}
\end{table}

The index of refraction differs between Al$_2$O$_3$ solids with different lattice structure.
For amorphous Al$_2$O$_3$ at short wavelengths, \cite{Suh2016} calculated the index of refraction based on experimental data obtained by \cite{Begemann1997}
for wavelengths between $7.8$ and $500~\mu$m.  For crystalline Al$_2$O$_3$, \cite{Harman1994} and \cite{Koike1995}  measured the index of
refraction of $\alpha$-Al$_2$O$_3$ and $\gamma$-Al$_2$O$_3$, respectively. \citeauthor{Koike1995} studied two samples of $\gamma$-Al$_2$O$_3$,
one commercially available and the other obtained as a combustion product. \cite{Edlou1993} measured the optical constants of an Al$_2$O$_3$ coating deposited
by reactive magnetron sputtering. The lattice structure of the coating was not specified by the authors. A summary of the relevant values obtained from the
literature is given in Table~\ref{tab:OptConst}.

The large spread in values of $\kappa$ reported for Al$_2$O$_3$ solids leads to very uncertain absorption properties of Al$_2$O$_3$ grains, while the small spread in the value of $n$
suggests some confidence in the value of the scattering properties. If the value of $\kappa$
is relatively too large, the equilibrium temperature of the grains might be too high
for them to remain amorphous, or even form at all, close to the star.
\cite{Hoefner2016} found that aluminium oxide dust can form very efficiently if the grains have low absorption cross sections (corresponding to low values
of $\kappa$). The authors found that if the value of $\kappa$ for aluminium oxide
grains is similar to that of Mg$_2$SiO$_4$ ($\sim 10^{-4}$),
aluminium would be fully depleted at $r \sim 2.1$~au (1.5~$R^{\rm IR}_\star$).
If the same applies to aluminium oxide grains around Mira~A, full condensation of aluminium could
happen already below the millimetre and visible stellar photosphere.

\subsubsection{Model calculations and results}

The grain opacities were calculated using the hollow-spheres approximation \citep{Min2003}. We used the complex refractive index
obtained by \cite{Suh2016} in our model calculations, since grains with these properties scatter less efficiently.
The scattering opacity we find at 0.82~$\mu$m using the data from \cite{Suh2016} is 
${\sim 25\%}$ smaller than that obtained using the optical constants provided by \cite{Koike1995}.
MCMax calculates the scattering of radiation considering the full angle-dependent Mueller matrix and produces the images of the Stokes parameters
$I$, $Q$, and $U$. We convolved these images with the observed PSF reference to obtain images
at the same resolution as  the observed ones ($I^{\rm conv}$, $Q^{\rm conv}$, and $U^{\rm conv}$).
Then, we constructed the images of the polarized intensity and the polarization degree using the convolved images and Eqs. \ref{equ:polInt} and \ref{equ:polDeg}.
We only model the observations at $0.82~\mu$m.

Our modelling focuses on the region traced by the vibrationally-excited CO line, with an outer radius of $5.6~R^{\rm IR}_\star = 4~R^{\rm 338~GHz}_\star =
8.6$~au. We consider a solar abundance of aluminium of $2.8 \times 10^{-6}$ relative to H \citep{Asplund2009}, which
implies the same value for the maximum abundance of Al$_2$O$_3$ relative to H$_2$. Based on the gas model
presented in Sect.~\ref{sec:modelCO}, the solar abundance of Al,
and the mass of an Al$_2$O$_3$ molecule ($1.7 \times 10^{-22}$~g),
we write the radial density profile of Al$_2$O$_3$ dust grains as
\begin{equation*}
\rho^{\rm Al_2O_3}_\circ(r) \approx  4.8 \times 10^{-28} ~n_\circ~ f~(1.4 \times R^{\rm IR}_\star/r)^\eta ~{\rm g~cm}^{-3},
\end{equation*}
for $R^{\rm Dust}_{\rm in} < r < R^{\rm Dust}_{\rm out}$. In this expression, $f$ is the fraction of aluminium that condenses into Al$_2$O$_3$ dust, $R^{\rm Dust}_{\rm in}$ is the inner radius of the dust envelope (see below),
and $R^{\rm Dust}_{\rm out} = 5.6~R^{\rm IR}_\star$ is the outer radius of all model dust envelopes. The values of $n_\circ$ and $\eta$ are adopted
from the gas model.

We calculated models { to fit the cnt820 images} assuming that condensation of aluminium happens at $r \sim 22.5~{\rm mas} = 1.5~R^{\rm IR}_\star$ or at
$r \sim 37.5~{\rm mas} = 2.5~R^{\rm IR}_\star$, based, respectively, on theoretical results and on interferometric mid-infrared observations of
other Mira variables, as discussed above.
The star was approximated by a black body with a temperature of 2500~K. This assumption does not strongly affect the calculations of the scattered light.
The free parameters in our dust models are $f$ and the size of the dust grains.
We consider an upper limit to the polarization degree of 2\% in the vibrationally-excited CO region (see discussion in Sect. \ref{sec:beam-shift}).

For both condensation radii considered,
our models show that only $< 10\%$, $< 1\%$, $< 1\%$, and $\sim 100\%$ of the available aluminium atoms
can be locked in $1.0~\mu$m, $0.3~\mu$m, $0.1~\mu$m, and $0.02~\mu$m aluminium oxide grains, respectively.
Otherwise, a stronger polarized-light signal
at visible wavelengths would be expected.
This result is valid for all the values of $n$ given in Table~\ref{tab:OptConst}, and is therefore independent of the assumed lattice structure.
Al$_2$O$_3$ grains with sizes of $0.1~\mu$m could account for $> 10$\% of the aluminium atoms in the vibrationally-excited CO region only if the value of $n$ of
such grains was $< 1.05$. This is significantly smaller than what the experimental results suggest for
Al$_2$O$_3$ solids with different lattice structures. We find that a relatively larger population ($\sim 10\%$) of grains with sizes $\sim 1~\mu$m could
exist, although the opacity per unit mass they provide is larger than that provided by $\sim 0.1~\mu$m grains.
The lower polarization degree is caused by the more complex scattering phase function
of the $1.0~\mu$m grains. We consider it unlikely that a population of grains of $\sim 1~\mu$m
 exists around Mira without a significant amount of grains with size between 0.02 and $0.5~\mu$m. Moreover, even if $\sim 1~\mu$m-sized grains do exist,
 these would still only account for a small fraction of the expected amount of aluminium atoms close to Mira~A.
We conclude that if aluminium atoms in the vibrationally-excited CO region
are efficiently locked in aluminium oxide grains, these must have relatively small sizes $\lesssim 0.02~\mu$m.
This is not consistent with the larger grains ($\sim 0.1~\mu$m) found in models for Al$_2$O$_3$ dust formation and growth
in the extended atmospheres of AGB stars.
{ We consider it unlikely that a significant fraction of aluminium atoms are locked in $\sim 0.1~\mu$m grains of
a dust species other than aluminium oxide (such as in a grain with an Al$_2$O$_3$ core and a silicate mantle). This is because
aluminium would account for a smaller fraction of the mass of a single grain in that case and, therefore, significant depletion of aluminium
would lead to a larger amount of dust. Such dust species could therefore only exist if they were very transparent at visible wavelengths.
Optical constants of Mg$_2$SiO$_4$ and MgSiO$_3$ \citep{Jager2003} indicate that these silicates
have scattering properties that are similar to those of Al$_2$O$_3$.}

\cite{Decin2017} suggested the bulk of aluminium atoms
could be locked in (Al$_2$O$_3$)$_n$ clusters, which would have sizes consistent with the limits we set.
These clusters
might still produce an infrared excess similar to that expected from amorphous aluminium oxide grains, and therefore be detected in
interferometric observations in the mid-infrared.
Although experiments have shown that large aluminium oxide clusters (with more than 34 Al$_2$O$_3$ molecules)
 have a similar spectral signature to that of amorphous aluminium oxide \citep{vanHeijnsbergen2003},
the opacity per mass of such clusters is not known.

\subsubsection{The effect of Al$_2$O$_3$ opacity at mid-infrared wavelengths}

Considering either the optical constants given by \cite{Koike1995} for crystalline aluminium oxide or \cite{Begemann1997} for
amorphous aluminium oxide, we find that the inner part of the envelope would become optically thick at the
aluminium oxide resonance at $\sim11~\mu$m for full aluminium condensation.
The opacity due to the efficient formation of large Al$_2$O$_3$ clusters cannot be determined based on the data available in the literature at the moment.
This would cause the size of Mira~A to be roughly two times larger at $11~\mu$m (45~mas in radius in our models) than at millimetre wavelengths.
\cite{Weiner2003} observed Mira~A between 1999 and 2001
with the Infrared Spatial Interferometer (ISI) at $\sim 11~\mu$m and found a stellar radius varying between 23~mas and 31~mas.
This is larger than the radius measured in the near-infrared by roughly a factor of two, and therefore comparable to the millimetre
and visible sizes we report { ($\sim 21$~mas)}.
Hence this measurement does not support the existence of a shell of aluminium-oxide dust that is optically thick at $\sim 11~\mu$m.
Models by \cite{Weiner2004} and \cite{Ohnaka2006} explained the observed 11~$\mu$m size of Mira~A with an optically-thick layer of hot (2200~K) water,
although some discrepancies between model and observations remained. \cite{Ohnaka2006} argued that dust{ formation on the upper atmosphere}
could also contribute significantly to the opacity at 11~$\mu$m.

More recent measurements with mid-infrared interferometric instrument (MIDI) show an asymmetric and relatively larger star
between 10 and 14~$\mu$m, with radii as large as $70$~mas (private communication Claudia Paladini).
This would be comparable to what we expect from an optically-thick aluminium
oxide shell. The complexity of the source at the time of the MIDI observations makes an objective conclusion not possible.
Moreover, the distribution of gas close to the star at the time of the ISI
and MIDI observations was most likely very different from that shown by the observations we report. 
New spatially-resolved observations at $\sim 11~\mu$m, possibly together with coordinated ALMA and SPHERE observations, will
help further constrain the presence and the properties of aluminium oxide grains in the close vicinity of Mira~A.

\section{Summary}

We imaged molecular line emission and polarized light, produced by {\bf scattering of radiation on} dust grains,
around the Mira { AB system} using ALMA and SPHERE/ZIMPOL. The images in polarized light
show dust around Mira~A, but also around Mira~B and a dust trail connecting the two stars. 
We find a region of high
gas densities around Mira~A, which seems to be
delimited by a steep density decline at its outer edge. The dust grains that produce the observed scattered light cluster at the edge of this region.
If the proposed density decline is caused by the propagation of a shock front,
this would be in qualitative agreement with models that predict dust formation to happen very efficiently in the post-shocked gas. The radius at 
which the density drop is seen ($\sim 12~R^{\rm IR}_\star$) is larger than the distances at which pulsation-induced shocks dissipate in hydrodynamical models, however. 
Alternative possibilities are that the mass-loss rate of Mira has recently increased creating an inner region with high gas densities, or that interactions
with the binary companion enhance the density in this region.
We model the CO~$v=1, J=3-2$ line{ to obtain}
the radial profiles of the gas density, temperature, and velocity very close to the star.
The gas model was refined by comparing different models to the $^{13}$CO~$J=3-2$ line data.
We calculate the abundances of TiO$_2$ and AlO and find that these account for only 4.5\% of Ti and $< 0.1\%$ of Al expected for
a gas of solar elemental abundance. The lack of TiO lines and observations of other potential Ti carriers (such as atomic Ti)
in our dataset makes a firm conclusion on the Ti depletion difficult, but
we estimate that TiO and TiO$_2$ account together for $\sim 20\%$ of the expected Ti.
It is even harder to determine the aluminium gas-phase budget, but we are able to constrain the properties and sizes of aluminium oxide dust grains
that might be an important reservoir of aluminium atoms.
We find that grains located within 84~mas ($5.6~R^{\rm IR}_\star$) from the star
must have an average size $\lesssim 0.02~\mu$m to account for a significant fraction of Al.
If Al$_2$O$_3$ grains with sizes $\sim 0.1~\mu$m are present, these must
account for less than 1\% of the aluminium atoms.
We also find that opacity due to efficient aluminium oxide dust formation could cause the star to appear very large
at wavelengths close to the $11~\mu$m feature. Recent interferometric observations in the mid-infrared suggest that Mira~A might indeed be very
large at these wavelengths, but
a detailed analysis is hampered by the
complexity of the source and the time lag between the mid-infrared and the ZIMPOL and ALMA observations that we report.
Quasi-simultaneous observations in visible scattered light, infrared thermal dust emission, and molecular line emission will help
better constrain the aluminium budget in Mira and other O-rich AGB stars.

\begin{acknowledgements}
TK, WV, HO, and MM  acknowledge  support  from  the  Swedish  Research  Council.
EDB acknowledges support from the Swedish National Space Agency. This paper
makes  use  of  the  following  ALMA  data:  ADS/JAO.ALMA\#2017.1.00191.S
and ADS/JAO.ALMA\#2012.1.00524.S.
ALMA is a partnership of ESO (representing its member states), NSF (USA)
and  NINS  (Japan),  together  with  NRC  (Canada),  NSC  and  ASIAA  (Tai-
wan),  and  KASI  (Republic  of  Korea),  in  cooperation  with  the  Republic
of  Chile.  The  Joint  ALMA  Observatory  is  operated  by  ESO,  AUI/NRAO
and NAOJ.
\end{acknowledgements}

\bibliographystyle{aa}
\bibliography{../../bibliography_2}

\begin{thebibliography}{54}
\expandafter\ifx\csname natexlab\endcsname\relax\def\natexlab#1{#1}\fi

\bibitem[{{Aronson} {et~al.}(2017){Aronson}, {Bladh}, \&
  {H{\"o}fner}}]{Aronson2017}
{Aronson}, E., {Bladh}, S., \& {H{\"o}fner}, S. 2017, \aap, 603, A116

\bibitem[{{Asplund} {et~al.}(2009){Asplund}, {Grevesse}, {Sauval}, \&
  {Scott}}]{Asplund2009}
{Asplund}, M., {Grevesse}, N., {Sauval}, A.~J., \& {Scott}, P. 2009, \araa, 47,
  481

\bibitem[{{Begemann} {et~al.}(1997){Begemann}, {Dorschner}, {Henning},
  {Mutschke}, {Guertler}, {Koempe}, \& {Nass}}]{Begemann1997}
{Begemann}, B., {Dorschner}, J., {Henning}, T., {et~al.} 1997, \apj, 476, 199

\bibitem[{{Bladh} \& {H{\"o}fner}(2012)}]{Bladh2012}
{Bladh}, S. \& {H{\"o}fner}, S. 2012, \aap, 546, A76

\bibitem[{{Bourges} {et~al.}(2017){Bourges}, {Mella}, {Lafrasse}, {Duvert},
  {Chelli}, {Le Bouquin}, {Delfosse}, \& {Chesneau}}]{Bourges2017}
{Bourges}, L., {Mella}, G., {Lafrasse}, S., {et~al.} 2017, VizieR Online Data
  Catalog, 2346

\bibitem[{{Decin} {et~al.}(2017){Decin}, {Richards}, {Waters}, {Danilovich},
  {Gobrecht}, {Khouri}, {Homan}, {Bakker}, {Van de Sande}, {Nuth}, \& {De
  Beck}}]{Decin2017}
{Decin}, L., {Richards}, A.~M.~S., {Waters}, L.~B.~F.~M., {et~al.} 2017, \aap,
  608, A55

\bibitem[{{Edlou} {et~al.}(1993){Edlou}, {Smajkiewicz}, \&
  {Al-Jumaily}}]{Edlou1993}
{Edlou}, S.~M., {Smajkiewicz}, A., \& {Al-Jumaily}, G.~A. 1993, \ao, 32, 5601

\bibitem[{{Freytag} \& {H{\"o}fner}(2008)}]{Freytag2008}
{Freytag}, B. \& {H{\"o}fner}, S. 2008, \aap, 483, 571

\bibitem[{{Freytag} {et~al.}(2017){Freytag}, {Liljegren}, \&
  {H{\"o}fner}}]{Freytag2017}
{Freytag}, B., {Liljegren}, S., \& {H{\"o}fner}, S. 2017, \aap, 600, A137

\bibitem[{{Gail} {et~al.}(2016){Gail}, {Scholz}, \& {Pucci}}]{Gail2016}
{Gail}, H.-P., {Scholz}, M., \& {Pucci}, A. 2016, \aap, 591, A17

\bibitem[{{Gendriesch} {et~al.}(2009){Gendriesch}, {Lewen}, {Klapper},
  {Menten}, {Winnewisser}, {Coxon}, \& {M{\"u}ller}}]{Gendriesch2009}
{Gendriesch}, R., {Lewen}, F., {Klapper}, G., {et~al.} 2009, \aap, 497, 927

\bibitem[{{Gobrecht} {et~al.}(2016){Gobrecht}, {Cherchneff}, {Sarangi},
  {Plane}, \& {Bromley}}]{Gobrecht2016}
{Gobrecht}, D., {Cherchneff}, I., {Sarangi}, A., {Plane}, J.~M.~C., \&
  {Bromley}, S.~T. 2016, \aap, 585, A6

\bibitem[{{Haniff} {et~al.}(1995){Haniff}, {Scholz}, \& {Tuthill}}]{Haniff1995}
{Haniff}, C.~A., {Scholz}, M., \& {Tuthill}, P.~G. 1995, \mnras, 276, 640

\bibitem[{{Harman} {et~al.}(1994){Harman}, {Ninomiya}, \&
  {Adachi}}]{Harman1994}
{Harman}, A.~K., {Ninomiya}, S., \& {Adachi}, S. 1994, Journal of Applied
  Physics, 76, 8032

\bibitem[{{Hinkle} {et~al.}(2016){Hinkle}, {Lebzelter}, \&
  {Straniero}}]{Hinkle2016}
{Hinkle}, K.~H., {Lebzelter}, T., \& {Straniero}, O. 2016, \apj, 825, 38

\bibitem[{{H{\"o}fner}(2008)}]{Hofner2008}
{H{\"o}fner}, S. 2008, \aap, 491, L1

\bibitem[{{H{\"o}fner} {et~al.}(2016){H{\"o}fner}, {Bladh}, {Aringer}, \&
  {Ahuja}}]{Hoefner2016}
{H{\"o}fner}, S., {Bladh}, S., {Aringer}, B., \& {Ahuja}, R. 2016, \aap, 594,
  A108

\bibitem[{{H{\"o}fner} {et~al.}(2003){H{\"o}fner}, {Gautschy-Loidl}, {Aringer},
  \& {J{\o}rgensen}}]{Hofner2003}
{H{\"o}fner}, S., {Gautschy-Loidl}, R., {Aringer}, B., \& {J{\o}rgensen}, U.~G.
  2003, \aap, 399, 589

\bibitem[{{H{\"o}fner} \& {Olofsson}(2018)}]{Hoefner2018}
{H{\"o}fner}, S. \& {Olofsson}, H. 2018, \aapr, 26, 1

\bibitem[{{Ireland} {et~al.}(2007){Ireland}, {Monnier}, {Tuthill}, {Cohen}, {De
  Buizer}, {Packham}, {Ciardi}, {Hayward}, \& {Lloyd}}]{Ireland2007}
{Ireland}, M.~J., {Monnier}, J.~D., {Tuthill}, P.~G., {et~al.} 2007, \apj, 662,
  651

\bibitem[{{Ireland} {et~al.}(2011){Ireland}, {Scholz}, \& {Wood}}]{Ireland2011}
{Ireland}, M.~J., {Scholz}, M., \& {Wood}, P.~R. 2011, \mnras, 418, 114

\bibitem[{{J{\"a}ger} {et~al.}(2003){J{\"a}ger}, {Dorschner}, {Mutschke},
  {Posch}, \& {Henning}}]{Jager2003}
{J{\"a}ger}, C., {Dorschner}, J., {Mutschke}, H., {Posch}, T., \& {Henning}, T.
  2003, \aap, 408, 193

\bibitem[{{Kami{\'n}ski} {et~al.}(2017){Kami{\'n}ski}, {M{\"u}ller}, {Schmidt},
  {Cherchneff}, {Wong}, {Br{\"u}nken}, {Menten}, {Winters}, {Gottlieb}, \&
  {Patel}}]{Kaminski2017}
{Kami{\'n}ski}, T., {M{\"u}ller}, H.~S.~P., {Schmidt}, M.~R., {et~al.} 2017,
  \aap, 599, A59

\bibitem[{{Kami{\'n}ski} {et~al.}(2016){Kami{\'n}ski}, {Wong}, {Schmidt},
  {M{\"u}ller}, {Gottlieb}, {Cherchneff}, {Menten}, {Keller}, {Br{\"u}nken},
  {Winters}, \& {Patel}}]{Kaminski2016}
{Kami{\'n}ski}, T., {Wong}, K.~T., {Schmidt}, M.~R., {et~al.} 2016, \aap, 592,
  A42

\bibitem[{{Kania} {et~al.}(2011){Kania}, {Hermanns}, {Br{\"u}nken},
  {M{\"u}ller}, \& {Giesen}}]{Kania2011}
{Kania}, P., {Hermanns}, M., {Br{\"u}nken}, S., {M{\"u}ller}, H.~S.~P., \&
  {Giesen}, T.~F. 2011, Journal of Molecular Spectroscopy, 268, 173

\bibitem[{{Karovicova} {et~al.}(2013){Karovicova}, {Wittkowski}, {Ohnaka},
  {Boboltz}, {Fossat}, \& {Scholz}}]{Karovicova2013}
{Karovicova}, I., {Wittkowski}, M., {Ohnaka}, K., {et~al.} 2013, \aap, 560, A75

\bibitem[{{Karovska} {et~al.}(2005){Karovska}, {Schlegel}, {Hack}, {Raymond},
  \& {Wood}}]{Karovska2005}
{Karovska}, M., {Schlegel}, E., {Hack}, W., {Raymond}, J.~C., \& {Wood}, B.~E.
  2005, \apjl, 623, L137

\bibitem[{{Kervella} {et~al.}(2018){Kervella}, {Decin}, {Richards}, {Harper},
  {McDonald}, {O'Gorman}, {Montarg{\`e}s}, {Homan}, \& {Ohnaka}}]{Kervella2018}
{Kervella}, P., {Decin}, L., {Richards}, A.~M.~S., {et~al.} 2018, \aap, 609,
  A67

\bibitem[{{Khouri} {et~al.}(2016{\natexlab{a}}){Khouri}, {Maercker}, {Waters},
  {Vlemmings}, {Kervella}, {de Koter}, {Ginski}, {De Beck}, {Decin}, {Min},
  {Dominik}, {O'Gorman}, {Schmid}, {Lombaert}, \& {Lagadec}}]{Khouri2016}
{Khouri}, T., {Maercker}, M., {Waters}, L.~B.~F.~M., {et~al.}
  2016{\natexlab{a}}, \aap, 591, A70

\bibitem[{{Khouri} {et~al.}(2016{\natexlab{b}}){Khouri}, {Vlemmings},
  {Ramstedt}, {Lombaert}, {Maercker}, \& {De Beck}}]{Khouri2016b}
{Khouri}, T., {Vlemmings}, W.~H.~T., {Ramstedt}, S., {et~al.}
  2016{\natexlab{b}}, \mnras, 463, L74

\bibitem[{{Khouri} {et~al.}(2015){Khouri}, {Waters}, {de Koter}, {Decin},
  {Min}, {de Vries}, {Lombaert}, \& {Cox}}]{Khouri2015}
{Khouri}, T., {Waters}, L.~B.~F.~M., {de Koter}, A., {et~al.} 2015, \aap, 577,
  A114

\bibitem[{{Koike} {et~al.}(1995){Koike}, {Kaito}, {Yamamoto}, {Shibai},
  {Kimura}, \& {Suto}}]{Koike1995}
{Koike}, C., {Kaito}, C., {Yamamoto}, T., {et~al.} 1995, \icarus, 114, 203

\bibitem[{{Mart{\'{\i}}-Vidal} {et~al.}(2014){Mart{\'{\i}}-Vidal}, {Vlemmings},
  {Muller}, \& {Casey}}]{Marti-Vidal2014}
{Mart{\'{\i}}-Vidal}, I., {Vlemmings}, W.~H.~T., {Muller}, S., \& {Casey}, S.
  2014, \aap, 563, A136

\bibitem[{{Min} {et~al.}(2009){Min}, {Dullemond}, {Dominik}, {de Koter}, \&
  {Hovenier}}]{Min2009}
{Min}, M., {Dullemond}, C.~P., {Dominik}, C., {de Koter}, A., \& {Hovenier},
  J.~W. 2009, \aap, 497, 155

\bibitem[{{Min} {et~al.}(2003){Min}, {Hovenier}, \& {de Koter}}]{Min2003}
{Min}, M., {Hovenier}, J.~W., \& {de Koter}, A. 2003, \aap, 404, 35

\bibitem[{{Mohamed} \& {Podsiadlowski}(2012)}]{Mohamed2012}
{Mohamed}, S. \& {Podsiadlowski}, P. 2012, Baltic Astronomy, 21, 88

\bibitem[{{Norris} {et~al.}(2012){Norris}, {Tuthill}, {Ireland}, {Lacour},
  {Zijlstra}, {Lykou}, {Evans}, {Stewart}, \& {Bedding}}]{Norris2012}
{Norris}, B.~R.~M., {Tuthill}, P.~G., {Ireland}, M.~J., {et~al.} 2012, \nat,
  484, 220

\bibitem[{{Ohnaka} {et~al.}(2006){Ohnaka}, {Scholz}, \& {Wood}}]{Ohnaka2006}
{Ohnaka}, K., {Scholz}, M., \& {Wood}, P.~R. 2006, \aap, 446, 1119

\bibitem[{{Ohnaka} {et~al.}(2016){Ohnaka}, {Weigelt}, \&
  {Hofmann}}]{Ohnaka2016}
{Ohnaka}, K., {Weigelt}, G., \& {Hofmann}, K.-H. 2016, \aap, 589, A91

\bibitem[{{Ohnaka} {et~al.}(2017){Ohnaka}, {Weigelt}, \&
  {Hofmann}}]{Ohnaka2017}
{Ohnaka}, K., {Weigelt}, G., \& {Hofmann}, K.-H. 2017, \aap, 597, A20

\bibitem[{{Ramstedt} {et~al.}(2014){Ramstedt}, {Mohamed}, {Vlemmings},
  {Maercker}, {Montez}, {Baudry}, {De Beck}, {Lindqvist}, {Olofsson},
  {Humphreys}, {Jorissen}, {Kerschbaum}, {Mayer}, {Wittkowski}, {Cox},
  {Lagadec}, {Leal-Ferreira}, {Paladini}, {P{\'e}rez-S{\'a}nchez}, \&
  {Sacuto}}]{Ramstedt2014}
{Ramstedt}, S., {Mohamed}, S., {Vlemmings}, W.~H.~T., {et~al.} 2014, \aap, 570,
  L14

\bibitem[{{Schmid} {et~al.}(2017){Schmid}, {Bazzon}, {Milli}, {Roelfsema},
  {Engler}, {Mouillet}, {Lagadec}, {Sissa}, {Sauvage}, {Ginski}, {Baruffolo},
  {Beuzit}, {Boccaletti}, {Bohn}, {Claudi}, {Costille}, {Desidera}, {Dohlen},
  {Dominik}, {Feldt}, {Fusco}, {Gisler}, {Girard}, {Gratton}, {Henning},
  {Hubin}, {Joos}, {Kasper}, {Langlois}, {Pavlov}, {Pragt}, {Puget}, {Quanz},
  {Salasnich}, {Siebenmorgen}, {Stute}, {Suarez}, {Szul{\'a}gyi}, {Thalmann},
  {Turatto}, {Udry}, {Vigan}, \& {Wildi}}]{Schmid2017}
{Schmid}, H.~M., {Bazzon}, A., {Milli}, J., {et~al.} 2017, \aap, 602, A53

\bibitem[{{Schmid} {et~al.}(2018){Schmid}, {Bazzon}, {Roelfsema}, {Mouillet},
  {Milli}, {Menard}, {Gisler}, {Hunziker}, {Pragt}, {Dominik}, {Boccaletti},
  {Ginski}, {Abe}, {Antoniucci}, {Avenhaus}, {Baruffolo}, {Baudoz}, {Beuzit},
  {Carbillet}, {Chauvin}, {Claudi}, {Costille}, {Daban}, {de Haan}, {Desidera},
  {Dohlen}, {Downing}, {Elswijk}, {Engler}, {Feldt}, {Fusco}, {Girard},
  {Gratton}, {Hanenburg}, {Henning}, {Hubin}, {Joos}, {Kasper}, {Keller},
  {Langlois}, {Lagadec}, {Martinez}, {Mulder}, {Pavlov}, {Podio}, {Puget},
  {Quanz}, {Rigal}, {Salasnich}, {Sauvage}, {Schuil}, {Siebenmorgen}, {Sissa},
  {Snik}, {Suarez}, {Thalmann}, {Turatto}, {Udry}, {van Duin}, {van Holstein},
  {Vigan}, \& {Wildi}}]{Schmid2018}
{Schmid}, H.~M., {Bazzon}, A., {Roelfsema}, R., {et~al.} 2018, ArXiv e-prints

\bibitem[{{Suh}(2016)}]{Suh2016}
{Suh}, K.-W. 2016, Journal of Korean Astronomical Society, 49, 127

\bibitem[{{van Heijnsbergen} {et~al.}(2003){van Heijnsbergen}, {Demyk},
  {Duncan}, {Meijer}, \& {von Helden}}]{vanHeijnsbergen2003}
{van Heijnsbergen}, D., {Demyk}, K., {Duncan}, M.~A., {Meijer}, G., \& {von
  Helden}, G. 2003, Physical Chemistry Chemical Physics (Incorporating Faraday
  Transactions), 5, 2515

\bibitem[{{van Leeuwen}(2007)}]{vanLeeuwen2007}
{van Leeuwen}, F. 2007, \aap, 474, 653

\bibitem[{{Vlemmings} {et~al.}(2017{\natexlab{a}}){Vlemmings}, {Khouri},
  {O'Gorman}, {De Beck}, {Humphreys}, {Lankhaar}, {Maercker}, {Olofsson},
  {Ramstedt}, {Tafoya}, \& {Takigawa}}]{Vlemmings2017}
{Vlemmings}, W., {Khouri}, T., {O'Gorman}, E., {et~al.} 2017{\natexlab{a}},
  Nature Astronomy, 1, 848

\bibitem[{{Vlemmings} {et~al.}(2017{\natexlab{b}}){Vlemmings}, {Khouri},
  {Mart{\'{\i}}-Vidal}, {Tafoya}, {Baudry}, {Etoka}, {Humphreys}, {Jones},
  {Kemball}, {O'Gorman}, {P{\'e}rez-S{\'a}nchez}, \&
  {Richards}}]{Vlemmings2017a}
{Vlemmings}, W.~H.~T., {Khouri}, T., {Mart{\'{\i}}-Vidal}, I., {et~al.}
  2017{\natexlab{b}}, \aap, 603, A92

\bibitem[{{Weiner}(2004)}]{Weiner2004}
{Weiner}, J. 2004, \apjl, 611, L37

\bibitem[{{Weiner} {et~al.}(2003){Weiner}, {Hale}, \& {Townes}}]{Weiner2003}
{Weiner}, J., {Hale}, D.~D.~S., \& {Townes}, C.~H. 2003, \apj, 588, 1064

\bibitem[{{Woitke}(2006)}]{Woitke2006}
{Woitke}, P. 2006, \aap, 460, L9

\bibitem[{{Wong} {et~al.}(2016){Wong}, {Kami{\'n}ski}, {Menten}, \&
  {Wyrowski}}]{Wong2016}
{Wong}, K.~T., {Kami{\'n}ski}, T., {Menten}, K.~M., \& {Wyrowski}, F. 2016,
  \aap, 590, A127

\bibitem[{{Woodruff} {et~al.}(2009){Woodruff}, {Ireland}, {Tuthill}, {Monnier},
  {Bedding}, {Danchi}, {Scholz}, {Townes}, \& {Wood}}]{Woodruff2009}
{Woodruff}, H.~C., {Ireland}, M.~J., {Tuthill}, P.~G., {et~al.} 2009, \apj,
  691, 1328

\bibitem[{{Zhao-Geisler} {et~al.}(2012){Zhao-Geisler}, {Quirrenbach},
  {K{\"o}hler}, \& {Lopez}}]{Zhao-Geisler2012}
{Zhao-Geisler}, R., {Quirrenbach}, A., {K{\"o}hler}, R., \& {Lopez}, B. 2012,
  \aap, 545, A56

\end{thebibliography}

\section*{Appendix}

\begin{figure*}[t]
   \centering
      \includegraphics[width= 16cm]{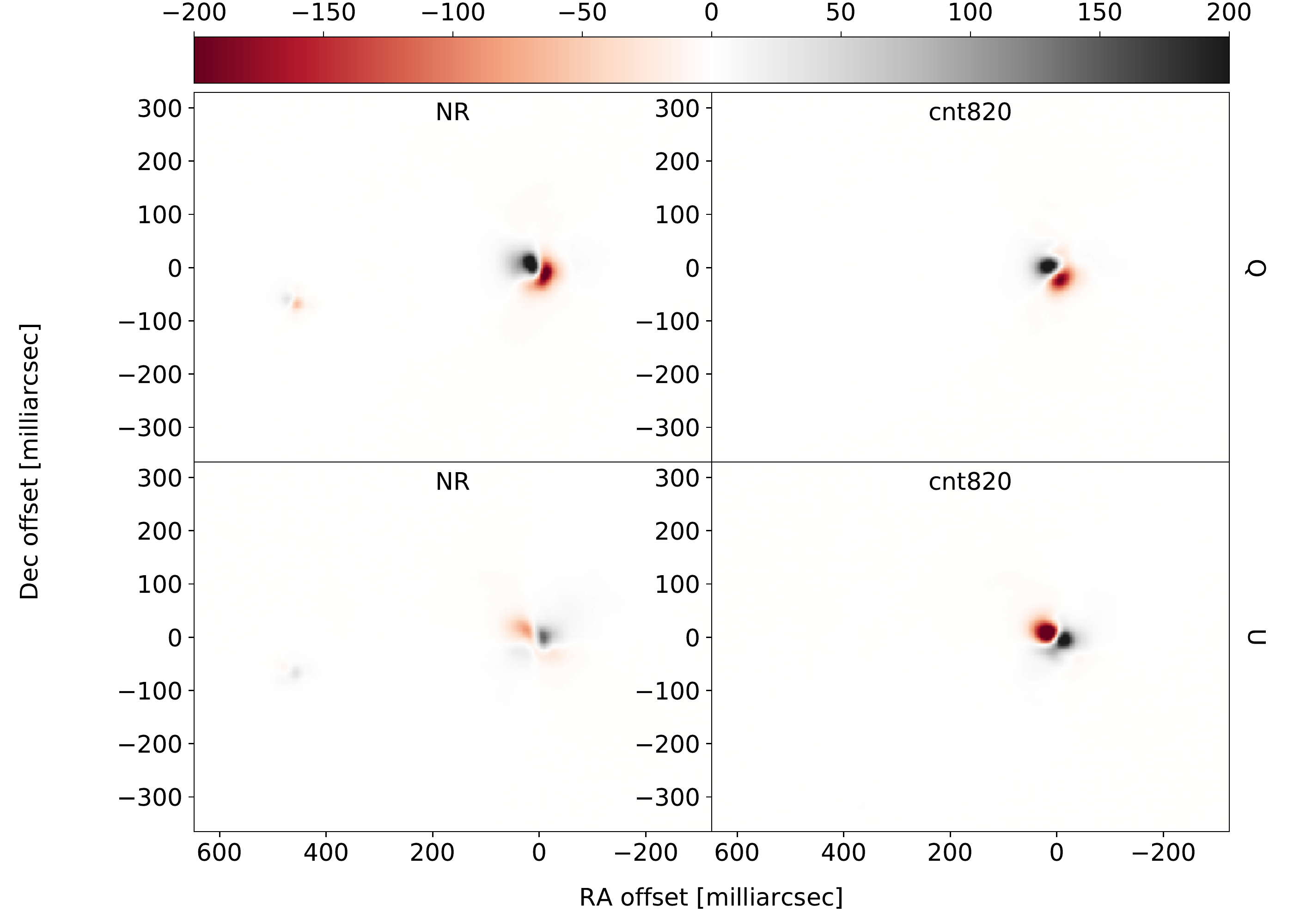}
      \hspace{0.5cm}
      \includegraphics[width= 16cm]{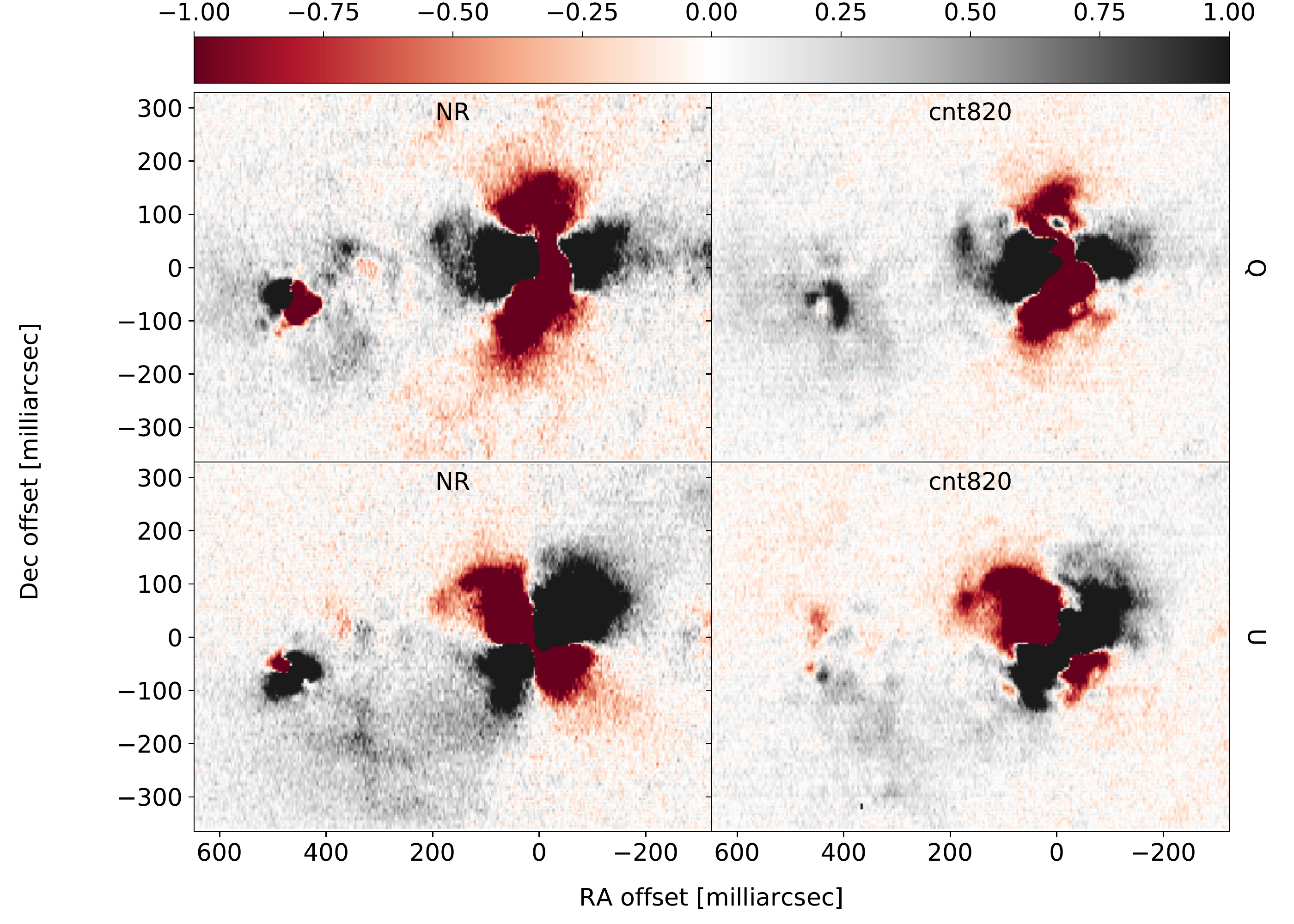}
      \caption{Images of Stokes parameters Q and U in filters NR and cnt820. { The colour maps show the observed counts toward the Mira~AB system and their 
      range differs to} highlight the beam-shift effect (upper panels)
      and the polarized-light signal (lower panels). { The offsets in declination and right ascension are given with respect to the position of Mira~A.}}
         \label{fig:Q+U_images}
   \end{figure*}

 \begin{figure*}[t]
   \centering
      \includegraphics[width= 18cm]{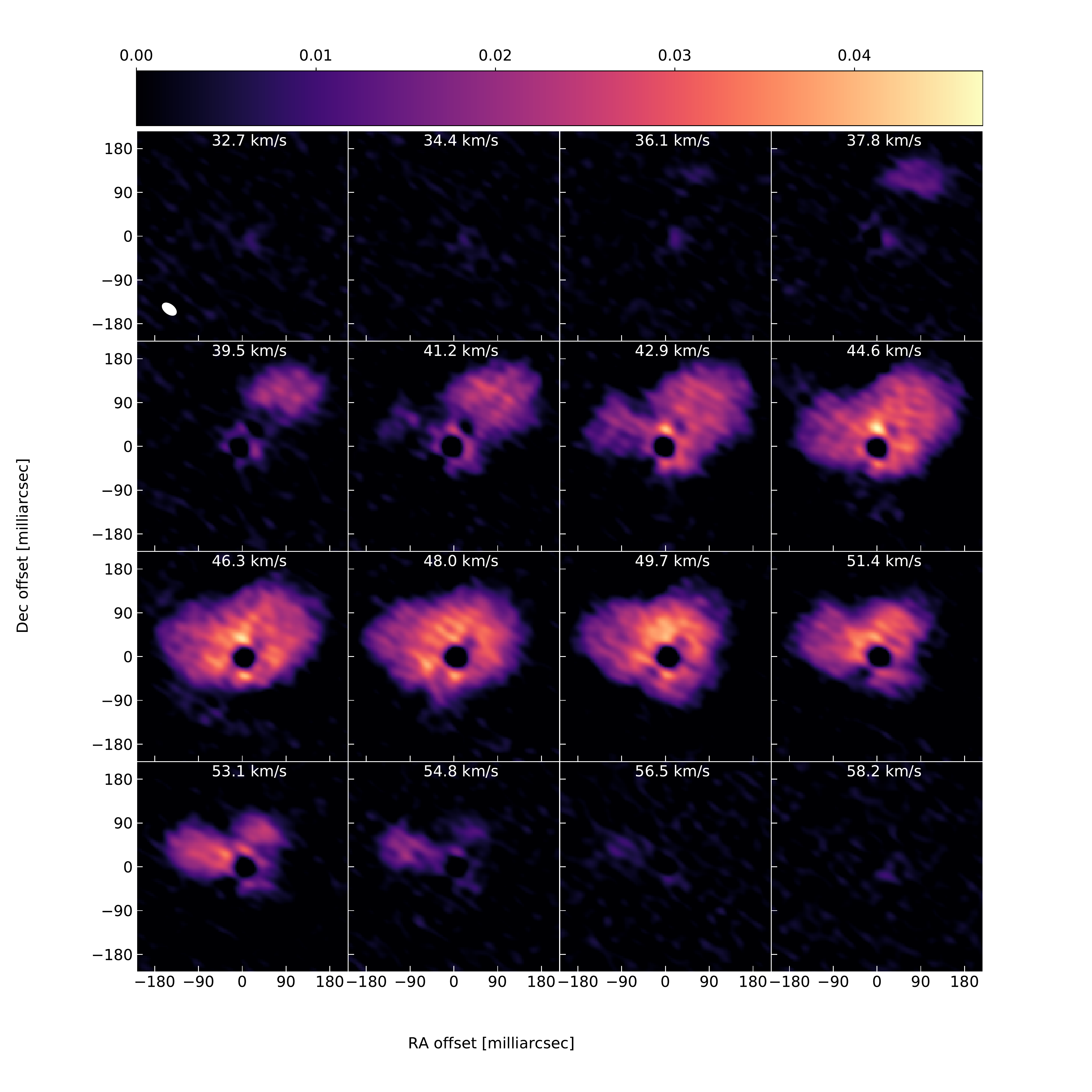}
      \caption{Velocity-channel maps of the SO~$N_J=8_8 - 7_7$ transition at 344.253~GHz given in Jy/beam. Each map is the average of emission within a channel $1.9$~km/s wide centred on the given local-standard-of-rest velocity. { The offsets in declination and right ascension are given with respect to the position of Mira~A.}}
         \label{fig:SO-channels}
   \end{figure*}

\end{document}